\documentclass[11pt,onecolumn,draftcls,journal]{IEEEtran}
\usepackage{amsfonts}
\usepackage{amssymb}
\usepackage{cite}
\ifCLASSINFOpdf
  \usepackage[pdftex]{graphicx}
\else
  \usepackage[dvips]{graphicx}
\fi
\usepackage[cmex10]{amsmath}
\usepackage{array}
\usepackage[tight,footnotesize]{subfigure}
\usepackage{caption}
\begin{document}

\title{Soft-In Soft-Out DFE and Bi-directional DFE
}

%\author{Seongwook Jeong and Jaekyun Moon}
%\normalsize{}
%\maketitle

%\author{\IEEEauthorblockN{Seongwook Jeong}
%\IEEEauthorblockA{Dept. of Electrical and Computer Engineering\\
%University of Minnesota\\
%Minneapolis, Minnesota 55455\\
%Email: jeong030@umn.edu} \and \IEEEauthorblockN{Jaekyun Moon}

%\IEEEauthorblockA{Dept. of Electrical and Computer Engineering\\
%University of Minnesota\\
%Minneapolis, Minnesota 55455\\
%Email: moon@umn.edu}}



\author{\IEEEauthorblockN{Seongwook Jeong, \IEEEmembership{Student Member, IEEE}
and Jaekyun Moon\IEEEauthorrefmark{2}, \IEEEmembership{Fellow,
IEEE}\\}
\IEEEauthorblockA{Dept. of Electrical and Computer Engineering\\
University of Minnesota\\
Minneapolis, Minnesota 55455, U.S.A.\\
Email: jeong030@umn.edu\\}
\IEEEauthorblockA{\IEEEauthorrefmark{2} Dept. of Electrical Engineering\\
Korea Advanced Institute of Science and Technology\\
Daejeon, 305-701, Republic of Korea \\
Email: jmoon@kaist.edu}

\thanks{This work was supported in part by the National Research Foundation of Korea under grant no. 2010-0029205 and the NSF under Theoretical Foundation grant no. 0728676 and IHCS grant no. 0701946. The material in this paper was presented in part at ICC 2010, Cape Town, South Africa, May 2010.}
}

\markboth{To appear in IEEE Transactions on Communications}{Jeong \MakeLowercase{and} Moon: Soft-in Soft-out DFE and BiDFE}


\maketitle
\setlength\arraycolsep{1pt}
\thispagestyle{empty}

\begin{abstract}
%\boldmath
We design a soft-in soft-out (SISO) decision feedback equalizer
(DFE) that performs better than its linear counterpart in turbo
equalizer (TE) setting. Unlike previously developed SISO-DFEs, the
present DFE scheme relies on extrinsic information formulation
that directly takes into account the error propagation effect.
With this new approach, both error rate simulation and the
extrinsic information transfer (EXIT) chart analysis indicate that
the proposed SISO-DFE is superior to the well-known SISO linear
equalizer (LE). This result is in contrast with the general
understanding today that the error propagation effect of the DFE
degrades the overall TE performance below that of the TE based on
a LE. We also describe a new extrinsic information combining
strategy involving the outputs of two DFEs running in opposite
directions, that explores error correlation between the two sets
of DFE outputs. When this method is combined with the new DFE
extrinsic information formulation, the resulting ``bidirectional"
turbo-DFE achieves excellent performance-complexity tradeoffs
compared to the TE based on the BCJR algorithm or on the LE.
Unlike turbo LE or turbo DFE, the turbo BiDFE's performance does
not degrade significantly as the feedforward and feedback filter
taps are constrained to be time-invariant.
\end{abstract}

% \begin{center} \bfseries Appear in Trans. Communications \end{center}
% \fi
%\IEEEpeerreviewmaketitle

%\newpage
\newpage
\setcounter{page}{1}
\section{Introduction}
Intersymbol interference (ISI) arises as the transmitted symbols
overlaps with one another in high speed digital communication.
Powerful modern equalization methods are based on the turbo
equalization principle established in \cite{Turbo}, wherein a
soft-in soft-out (SISO) equalizer (or detector) and a SISO
error-correction decoder exchange soft information in an iterative
fashion until reliable decisions are generated. It has been shown
in \cite{Turbo} that even for some heavy ISI channels the
detrimental effect of ISI disappears with this approach.

The detector or the equalizer portion of a turbo equalizer (TE)
system often investigated is based on the well-known
Bahl-Cocke-Jelinek-Raviv (BCJR) algorithm \cite{BCJR74}. This
algorithm exactly computes the \textit{a posteriori} probability
(APP) of the transmitted signal symbols considering the channel
response and the \textit{a priori} information of the transmitted
symbols and, as such, can be viewed as an optimum SISO equalizer.
However, the computational complexity of this algorithm grows
exponentially as a function of the channel length and the symbol
alphabet set size.

The high computational complexity of the BCJR-based equalizer has
motivated considerable research on numerous suboptimal but low
complexity equalization schemes. A notable development along this
direction is the well-known SISO linear equalizer (LE) of
\cite{TE02}. Another possibility, which was also evaluated in
\cite{TE02}, is the SISO decision feedback equalizer (DFE). In the
classical, non-turbo setting (i.e., no iterative exchange of soft
information between the equalizer and the decoder), it has long
been known that the DFE almost always outperforms the LE, despite
the fact that the DFE typically suffers from error propagation.
This is because when ISI is severe with the channel response
showing nulls or deep valleys within the Nyquist band, the LE is
subject to large noise enhancement. The work of \cite{TE02},
however, shows that when hard decisions are fed through the
feedback filter (to reduce complexity), SISO-DFE performs
considerably worse than SISO-LE, presumably due to error
propagation.

In classical DFE setting, many techniques have been investigated
to mitigate error propagation \cite{Fert98}, \cite{Ari98},
\cite{SFE06}. Recently, it has been shown \cite{BiDFE00},
\cite{BAD05}, \cite{BiSFE06} that conducting both normal and
time-reversed equalization of the received data sequence with two
DFEs running in opposite directions and combining two DFE outputs
is very effective in reducing error propagation and improving bit
error rate (BER) performance. This ``bi-directional" DFE (called
BiDFE) algorithm takes advantage of the different decision error
and noise distributions at the outputs of the forward and
time-reversed DFEs \cite{BiDFE00}, \cite{BAD05}.

The contribution of this paper is two-fold. One is that this paper
readdresses the DFE design issue in the turbo equalizer
environment and shows that just as in classical non-turbo setting,
the DFE outperforms the LE, if extrinsic information is
reformulated in a way that combats error propagation more
effectively. The second contribution is a specific DFE extrinsic
information combining strategy applied to a BiDFE that suppresses
statistical correlation between the outputs of two opposite
direction DFEs. We show that the resulting turbo BiDFE performance
approaches the performance of the BCJR-based turbo equalizer in a
fairly severe ISI environment, easily outperforming the turbo
equalizer based on the SISO-LE of \cite{TE02}. Remarkably, the
performance of a time-invariant version of the BiDFE, a
lower-complexity method that does not require tap-weight updating
as a function of time, also consistently is better than the
SISO-LE scheme of \cite{TE02} based on a time-varying linear
filter. There also exist feedback equalization techniques that
utilize soft decisions to reduce error propagation \cite{SFE06},
\cite{BiSFE06}, \cite{Jae05}, \cite{Rad05} but we focus on
hard-decision feedback in this paper, as the feedback
finite-impulse-response filter complexity is greatly reduced when
feedback decisions are constrained to take hard values.

The remainder of the paper is organized as follows. In Section
\ref{System_Model}, a brief statement of the problem is given. In
Section \ref{M-DFE}, we give a quick review of the SISO equalizer
design method established in \cite{TE02} and then provide a new
formulation of the extrinsic information of DFE taking into
account the error propagation effect. We also provide the
mean-squared-error analysis of the infinite-length BiDFE in
Section \ref{analysis1}. The iterative BiDFE algorithm is
introduced with the extrinsic information combiner of the normal
forward and time-reversed DFE outputs in Section \ref{BiDFEs}. In
Section \ref{simulation results}, numerical results and analysis
are given. Finally, we draw conclusions in Section
\ref{conclusion}.

\section{System Model}\label{System_Model}
We assume that the receiver knows the discrete-time baseband
channel response accurately. While the methods discussed are
general, our presentation will be based on binary symbols with
$P_x \triangleq \mathrm{E}(x_n^2) = 1$, $x_n \in \{ \pm 1 \}$, as
well as real-valued ISI channel coefficients and noise samples.
Although $x_n$ typically represents a coded bit sequence, our
analysis will assume that it is equiprobable and independent and
identically distributed (i.i.d.). Given the transmitted bit
sequence $\{x_k\}$, the channel output at time $n$ is
\begin{eqnarray} r_n & = & \sum_{k=0}^{L_h-1}
{h_k x_{n-k}} + w_n \label{eq:r_n}
\end{eqnarray}
where $w_n$ is additive white Gaussian
noise (AWGN) with variance $N_0$ and $\{h_k\}$ is
the channel impulse response with length $L_h$.

In turbo equalization, the equalizer computes the \textit{a
posteriori} log-likelihood ratio (LLR) of $x_n$,
\begin{eqnarray}
L(x_n) \triangleq \ln \dfrac{\mathrm{Pr}(x_n = +1 \mid \mathbf{r}_n)}{\mathrm{Pr}(x_n = -1 \mid \mathbf{r}_n)} \nonumber
\end{eqnarray}
where $\mathbf{r}_n$ is the received sample block utilized for LLR
estimation for $x_n$. Note that this computation requires the
knowledge of the \textit{a priori} probabilities of all input bits
affecting $\mathbf{r}_n$. Since these \textit{a priori}
probabilities are not available, they are all set to 1/2 initially
and then, as the turbo iteration ensues, to the estimated
probability values based on the extrinsic information generated
and passed back by the outer decoder.

The equalizer then generates its own extrinsic information by
subtracting the effect of the probability estimate passed down for
the current bit. Write this estimated \textit{a priori} LLR passed
down from the decoder as
\begin{eqnarray}
L_a(x_n) \triangleq \ln \dfrac{\mathrm{Pr}(x_n = +1)}{\mathrm{Pr}(x_n = -1)} \nonumber
\end{eqnarray}
with an understanding that the probabilities in the expression are
in reality just estimates.

Then, the equalizer's extrinsic LLR for $x_n$ to be passed to the
error-correction code decoder is given by
\begin{eqnarray}
L_e(x_n) \triangleq L(x_n) - L_a(x_n). \nonumber
\end{eqnarray}
This equation suggests first computing $L(x_n)$ based on the a
priori probabilities of all input bits including $x_n$ and then
simply subtracting $L_a(x_n)$ to generate the extrinsic LLR
$L_e(x_n)$. An alternative way of generating $L_e(x_n)$ is to set
$L_a(x_n)=0$ while computing $L(x_n)$, i.e., suppress the effect
of $L_a(x_n)$ in the calculation of $L(x_n)$:
\begin{eqnarray}
L_e(x_n) = L(x_n) \vert_{ L_a(x_n) = 0}. \nonumber
\end{eqnarray}
The techniques discussed in this paper actually use the second method.

\section{Derivation of Modified Iterative DFE Algorithm}\label{M-DFE}
In this section we first briefly review the results of \cite{TE02}
related to the SISO-DFE to provide necessary background while
establishing notation. We then show a new way of computing
extrinsic information so as to suppress error propagation and
improve performance.

\subsection{Review of Existing Extrinsic LLR Mapping}
The work of \cite{TE02} has established an effective strategy of
utilizing the \textit{a priori} information estimates from the
outer decoder in calculating the equalizer tap coefficients. The
gist of the approach in \cite{TE02} is a clever tweaking of the
classical minimum-mean-squared-error (MMSE) estimation principle
where the ``mean" of the input symbols are constructed using the
available \textit{a priori} information estimates and utilized in
the linear estimator weight computation. Both the LE and the DFE
can be designed in this way, but we shall focus on the DFE here.
Based on the above principle and suppressing the effect of the
\textit{a priori} probability estimate on the current bit $x_n$
(i.e., $\mathrm{E}(x_n)=0$) in an effort to extract the extrinsic
information, the MMSE feedforward filter taps (a total of $L_c+1$)
and the feedback filter taps (a total of $L_d=L_h-1$) at time $n$
are derived respectively as:
\begin{eqnarray}
\mathbf{c}_n  & \triangleq & \left[ {c_{\{n,0\}} ,c_{\{n,+ 1\}} , \ldots , c_{\{n,L_c\} } } \right]^T  \nonumber \\
& = & \left\{ {\mathbf{H}\mathbf{\Sigma}_n \mathbf{H}^T + (1 - z_n )\mathbf{s}\mathbf{s}^T + N_0 \mathbf{I}  } \right\}^{ - 1} \mathbf{s} \label{eq:c_n} \\
\mathbf{d}_n  & \triangleq & \left[ {d_{\{n,- L_d\} } ,d_{\{n,- L_d  + 1\}} , \ldots , d_{\{n,- 1\} }}  \right]^T  \nonumber \\
& = & \mathbf{M}\mathbf{H}^T \mathbf{c}_n \label{eq:d_n}
\end{eqnarray}
where $\mathbf{H}$ is a channel convolution matrix defined as
\begin{equation}
\mathbf{H} \triangleq \left [ {\begin{array}{*{20}c}
   {h_{L_h  - 1} } & {h_{L_h  - 2} } &  \cdots  & {h_0 } & 0 &  \cdots  & {} & 0  \\
   0 & {h_{L_h  - 1} } & {h_{L_h  - 2} } &  \cdots  & {h_0 } & 0 &  \cdots  & 0  \\
   {} & \ddots & {} &  \ddots  & {} & \ddots & {} & {}  \\
   0 & 0 &  \cdots  & 0 & {h_{L_h  - 1} } & {h_{L_h  - 2} } &  \cdots  & {h_0 }  \\
\end{array}} \right ] \nonumber
\end{equation}
and the matrix $\mathbf{\Sigma}_n$ depends on $\mathrm{E}(x_i)$,
 $i=n, n+1,..., n+L_c$, computed from the decoder output as
$\mathrm{E}(x_i) = \tanh(L_a(x_i)/2)$. Specifically,
$\mathbf{\Sigma}_n \triangleq {\mathrm{Diag}} ( \mathbf{0}_{1
\times L_d }, z_n, z_{n + 1}, \ldots, z_{n + L_c} )$ with $z_i
\triangleq 1 - [\mathrm{E}(x_i)]^2 $. Adding the term $(1 - z_n
)\mathbf{s}\mathbf{s}^T $ in (\ref{eq:c_n}) has the same effect of
suppressing $\mathrm{E}(x_n)$ to zero in
$\mathbf{H}\mathbf{\Sigma}_n \mathbf{H}^T$. The remaining vector
and matrix are defined as
 ${\mathbf{s}} \triangleq
{\mathbf{H}} [\mathbf{0}_{1 \times L_d } , 1 , \mathbf{0}_{1
\times L_c } ]^T$ and $\mathbf{M} \triangleq [\mathbf{I}_{L_d
\times L_d}, \mathbf{0}_{L_d \times (L_c+1)} ]$.

The equalizer output is obtained as
\begin{eqnarray}
y_n & = & \mathbf{c}_n^T \cdot \left( \mathbf{r}_n - \mathbf{H} \bar{\mathbf{x}}_n + \mathrm{E}(x_n) \mathbf{s} \right)
\label{eq:y_n_2}
\end{eqnarray}
where the received vector is defined as $\mathbf{r}_n \triangleq
\left[ {r_n ,r_{n + 1} , \ldots ,r_{n + L_c} } \right]^T$ and the
composite vector of the causal symbol decisions and the anticausal
symbols' mean as $\bar{\mathbf{x}}_n \triangleq \left[
\hat{x}_{n-L_d}, \ldots, \hat{x}_{n-1}, \mathrm{E}(x_{n}), \ldots,
\mathrm{E}(x_{n+L_c}) \right]^T$ where $\hat{x}_i$ is the
available decision for $x_i$ based on the \textit{a posteriori}
LLR of $x_i$, i.e., if $L(x_i)=L_a(x_i)+L_e(x_i) \geq 0$, then, $\hat{x}_i = +1$;
otherwise, $\hat{x}_i = -1$. The addition of the $\mathrm{E}(x_n)
\mathbf{s}$ term is also to suppress the effect of
$\mathrm{E}(x_n)$ in $\mathbf{H} \bar{\mathbf{x}}_n$.

Define the anticausal symbol sequence $\mathbf{x}_n \triangleq
\left[ {x_n ,x_{n + 1} , \ldots ,x_{n + L_c} }\right]^T$, the
causal symbol sequence $\mathbf{x}^c_n \triangleq \left[
{x_{n-L_d} ,x_{n - L_d+ 1} , \ldots ,x_{n -1}} \right]^T$, and the
available decision sequence $\mathbf{\hat{x}}^c_n \triangleq
\left[ {\hat{x}_{n-L_d}, \hat{x}_{n- L_d+ 1} , \ldots , \hat{x}_{n
-1} } \right]^T$. Also define the noise sequence as $\mathbf{w}_n
\triangleq \left[{w_n ,w_{n + 1} , \ldots ,w_{n + L_c} }
\right]^T$. Then, the combined filter output $y_n$ can be
rewritten as
\begin{eqnarray}
y_n & = & (\mathbf{c}^T_n \mathbf{H}_1) \cdot \big( \mathbf{x}_n - \mathrm{E}\{ \mathbf{\dot{x}}_n \} \big) + \mathbf{d}_n^T(\mathbf{x}^c_n - \mathbf{\hat{x}}^c_n) + \mathbf{c}^T_n\mathbf{w}_n \nonumber \\
& = & p_{\{n,0\}}x_n + \sum\limits_{k = 1}^{L_d} d_{\{n,-k\}}  \big( { x_{n - k}  - \hat{x}_{n - k} } \big) + \sum\limits_{k = 1}^{L_c} {p_{\{n,k\}} \big( {x_{n + k}  - \mathrm{E}(x_{n + k} )} \big)}  + \sum\limits_{k = 0}^{L_c} c_{\{n,k\}} w_{n+k} \nonumber \\
& = & p_{\{n,0\}}x_n + i_n + v_n \label{eq:new_model}
\end{eqnarray}
where $\mathrm{E}\{ \mathbf{\dot x}_n \} \triangleq \left[
{0,\mathrm{E}(x_{n + 1} ),\mathrm{E}(x_{n + 2} ),
\ldots,\mathrm{E}(x_{n + L_c} )} \right]^T$ and $\mathbf{H}_1$ is
the $(L_c+1) \times (L_c+1)$ submatrix of $\mathbf{H}$ formed by
the entire rows of the columns from the $(L_d+1)$th to the last.
Moreover, $\mathbf{p}_n \triangleq \left[ {p_{\{n,0\}}
,p_{\{n,1\}} , \ldots ,p_{\{n,L_c\}} } \right] = \mathbf{c}_n^T
\mathbf{H}_1$ and $p_{\{n,0\}} = \mathbf{c}^T_n \mathbf{s}$. The
error propagation caused by the mismatched hard decision feedback
is denoted as $i_n$, i.e., $i_n \triangleq  \sum_{k = 1}^{L_d}
d_{\{n,-k\}} \big( { x_{n - k}  - \hat{x}_{n - k} } \big)$ and
$v_n$ is the sum of noise and the remaining ISI terms caused by
the neighboring symbols: $v_n \triangleq \sum_{k = 1}^{L_c }
{p_{\{n,k\}} \big( {x_{n + k}  - \mathrm{E}(x_{n + k} )} \big)} +
\sum_{k = 0}^{L_c} c_{\{n,k\}} w_{n+k}$. The variance of $v_n$ is
\begin{eqnarray}
\mathrm{Var}(v_n) & \triangleq &  \mathbf{c}_n^T \mathrm{Cov} \{ \mathbf{r}_n \mathbf{r}_n^T \mid x_n = x \} \mathbf{c}_n \nonumber \\
& = & \mathbf{c}_n^T \mathbf{s} (1- \mathbf{s}^T \mathbf{c}_n).
\end{eqnarray}
Assuming that the feedback decisions are all correct, i.e., $i_n =
0$, and $v_n$ is AWGN, the extrinsic LLR is naturally given by
\begin{eqnarray}
L_e(x_n) &\triangleq & \ln \dfrac{\mathrm{Pr}(x_n = + 1 \mid y_n)}{\mathrm{Pr}(x_n = - 1 \mid y_n)} \Bigg\vert_{ L_a(x_n) = 0} \nonumber \\
& = & \ln \dfrac{\mathrm{Pr}(y_n  \mid x_n = + 1)\mathrm{Pr}(x_n = + 1)}{\mathrm{Pr}(y_n \mid x_n = - 1)\mathrm{Pr}(x_n = - 1)} \Bigg\vert_{ L_a(x_n) = 0} \nonumber \\
& = & \ln \dfrac{\mathrm{Pr}(y_n  \mid x_n = + 1)}{\mathrm{Pr}(y_n \mid x_n = - 1)} \nonumber \\
& = & - \dfrac{ \left | y_n - p_{\{n,0\}} \right |^2 }{2 \mathrm{Var} (v_n) }  +  \dfrac{ \left | y_n + p_{\{n,0\}}  \right |^2 }{2 \mathrm{Var} (v_n) }\nonumber \\
& = & \dfrac{2 p_{\{n,0\}}  y_n  }{\mathrm{Var} (v_n) }. \label{eq:Lx_n}
\end{eqnarray}
Notice that in generating $y_n$, $L_a(x_n)$ was already suppressed to zero.

A glossary of frequently used symbols is given below. Time-varying
quantities are augmented with time index $n$ as the subscript.

\begin{table}[h]
\renewcommand{\arraystretch}{1.0}
\centering

\begin{tabular}{|c|c||c|c|}
\hline
$\mathbf{c}_n$ & DFE feedforward filter coefficients of length $L_c+1$& $x_n$ & transmitted symbol \\
$\mathbf{d}_n$ & DFE feedback filter coefficients of length $L_d$ & $w_n$ & channel noise \\
$\mathbf{H}$ & channel convolution matrix & $P_x$ & average power of $x_n$\\
$\mathbf{M}$ & $[\mathbf{I}_{L_d \times L_d}, \mathbf{0}_{L_d \times (L_c+1)} ]$ & $N_0$ & variance of $w_n$\\
$\mathbf{s}$ & $\mathbf{H} [\mathbf{0}_{1 \times L_d } , 1 , \mathbf{0}_{1 \times L_c } ]^T$ & $\{h_k\}$ & ISI channel response of length $L_h$\\
$\mathbf{p}_n$ & $\mathbf{c}_n^T \mathbf{H}_1$ where $\mathbf{H}_1$ is a submatrix of $\mathbf{H}$ & $r_n$ & received channel output\\
$\mathbf{r}_n$ & received sample vector & $y_n$ & equalized observation\\
$\bar{\mathbf{x}}_n$ & vector of causal decisions and anticausal's mean & $i_n$ & error due to mismatched past decisions\\
$\mathbf{w}_n$ & noise sample vector & $v_n$ & noise plus error due to pre-cursor ISI\\
$\mathbf{x}_n$ & transmitted anticausal symbol vector & $p_{\{n,0\}}$ & weight on $x_n$ in $y_n$\\
$\mathbf{x}^c_n$ & transmitted causal symbol vector & $L_a(x_n)$ & \textit{a priori} LLR of $x_n$\\
$\mathbf{\hat{x}}^c_n$ & estimated causal symbol vector & $L(x_n)$ & \textit{a posteriori} LLR of $x_n$ \\
$\mathbf{y}^c_n$ & equalized causal sample vector & $L_e(x_n)$ & extrinsic LLR of $x_n$\\
$\mathbf{e}^c_{\{n,j\}}$ & possible causal error sequence & $z_n$ & variance of $x_n$\\
$\mathbf{\Sigma}_n$ & covariance matrix of transmitted anticausal symbols &  $\acute{z}_n$ & variance of $x_n$ estimated via \textit{a posteriori} LLR\\
$\mathbf{\acute{\Sigma}^c}_n$ & covariance matrix of estimated causal symbols & $\rho_n$ & noise correlation coefficient between two DFEs\\
\hline
\end{tabular}
\end{table}

\subsection{New Formulation of Extrinsic Information}\label{sec:M_Le}
While the MAP estimation of $i_n$ is equal to zero, we observe
that the chance of $i_n \neq 0$ is relatively high for severe ISI
channels. Our strategy is to estimate $i_n$ and utilize the
statistical parameters associated with this estimate in the
formulation of the extrinsic information. Since $i_n$ is to be
estimated on the basis of the observation $\mathbf{y}^c_n
\triangleq [y_{n-L_d}, y_{n-L_d+1}, \ldots, y_{n-1}]^T$, the mean
and variance of $i_n$ can be evaluated by the \textit{a
posteriori} probabilities of the causal symbols. Write
\begin{eqnarray}
\mathrm{E}(i_n) &\triangleq &  \mathrm{E} \left\{\mathbf{d}^T_n (\mathbf{x}^c_n - \mathbf{\hat{x}}^c_n) \mid \mathbf{y}^c_n \right\} \nonumber  \\
& = & \mathbf{d}_n^T \left( \tanh ( L(\mathbf{x}^c_n) /2) - \mathbf{\hat{x}}^c_n \right) \label{eq:mean(i_n)} \\
\mathrm{Var}(i_n) & \triangleq &  \mathrm{Var} \left\{ \mathbf{d}^T_n (\mathbf{x}^c_n - \mathbf{\hat{x}}^c_n) \mid \mathbf{y}^c_n \right \} \nonumber \\
& = & \mathbf{d}^T_n \mathbf{ \acute{\Sigma} }^c_n \mathbf{d}_n  \label{eq:sigma(i_n)^2}
\end{eqnarray}
where $L(\mathbf{x}^c_n) = [L(x_{n-L_d}), L(x_{n-L_d+1}), \ldots, L(x_{n-1})]^T$, $\mathbf{ \acute{\Sigma} }^c_n \triangleq \mathrm{Diag} \left( \acute{z}_{n-L_d}, \acute{z}_{n-L_d+1}, \ldots,
\acute{z}_{n-1} \right)$, and $\acute{z}_n = 1 - \tanh ( L(x_n) /2)^2$.

Now, let us consider the possible causal error sequence
$\mathbf{e}^c_{\{n,j\}} \triangleq \mathbf{x}^c_{\{n,j\}} -
\mathbf{\hat{x}}^c_n$ for $j=1,2,\ldots,2^{L_d}$, with index $j$
pointing to a particular binary pattern of $\mathbf{x}^c_n$. Then,
we can compute the extrinsic information for the given causal
error sequence $\mathbf{e}^c_{\{n,j\}}$:
\begin{eqnarray}
L_e(x_n | \mathbf{e}^c_{\{n,j\}} ) & \triangleq & \ln \dfrac{\mathrm{Pr}(y_n  \mid x_n = + 1, \mathbf{e}^c_{\{n,j\}})}{\mathrm{Pr}(y_n \mid x_n = - 1, \mathbf{e}^c_{\{n,j\}})} \nonumber \\
 & = & \dfrac{2 p_{\{n,0\}}  (y_n - \mathbf{d}_n^T \mathbf{e}^c_{\{n,j\}} ) }{\mathrm{Var} (v_n) }.\label{eq:Le/e}
\end{eqnarray}
To compute the extrinsic information of $x_n$ taking into account
the probabilities of possible error sequences, we write
\begin{eqnarray}
\mathrm{Pr}(y_n  \mid x_n = + 1) & =&  \sum_{j=1}^{2^{L_d}} \mathrm{Pr}(y_n  \mid x_n = +1, \mathbf{e}^c_{\{n,j\}}) \mathrm{Pr}(\mathbf{e}^c_{\{n,j\}}) \nonumber \\
& =& \sum_{j=1}^{2^{L_d}} \dfrac{\exp \left(L_e(x_n | \mathbf{e}^c_{\{n,j\}} ) \right)\mathrm{Pr}(\mathbf{e}^c_{\{n,j\}})}{1 + \exp \left( L_e(x_n | \mathbf{e}^c_{\{n,j\}} ) \right)}   \\
\mathrm{Pr}(y_n  \mid x_n = - 1) & = &\sum_{j=1}^{2^{L_d}} \mathrm{Pr}(y_n  \mid x_n = - 1, \mathbf{e}^c_{\{n,j\}}) \mathrm{Pr}(\mathbf{e}^c_{\{n,j\}}) \nonumber \\
& =& \sum_{j=1}^{2^{L_d}} \dfrac{\mathrm{Pr}(\mathbf{e}^c_{\{n,j\}})}{1 + \exp \left( L_e(x_n | \mathbf{e}^c_{\{n,j\}} ) \right )}.
\end{eqnarray}
Accordingly, the extrinsic information of $x_n$ considering the
distribution of $i_n$ is given as
\begin{eqnarray}
L_e(x_n) & = & \ln \left\{ \sum\limits_{j=1}^{2^{L_d}} \dfrac{ \exp \left( L_e(x_n | \mathbf{e}^c_{\{n,j\}} ) \right) \mathrm{Pr}(\mathbf{e}^c_{\{n,j\}} ) } { 1 + \exp \left( L_e(x_n | \mathbf{e}^c_{\{n,j\}} ) \right) } \right\} - \ln \left\{ \sum\limits_{j=1}^{2^{L_d}} \dfrac{ \mathrm{Pr}(\mathbf{e}^c_{\{n,j\}}) } { 1 + \exp \left( L_e(x_n | \mathbf{e}^c_{\{n,j\}} )  \right) } \right\} . \label{eq:Lx2_n}
\end{eqnarray}
In principle, the extrinsic information of (\ref{eq:Lx2_n}) can be
evaluated using (\ref{eq:Le/e}) and approximating
$\mathrm{Pr}(\mathbf{e}^c_{\{n,j\}})$ or
$\mathrm{Pr}(\mathbf{e}^c_{\{n,j\}} | \mathbf{y}_n^c)$ by
$\prod_{k=1}^{L_d} \mathrm{Pr}(e_{\{n-k,j\}} | y_{n-k}) $, which
can be computed based on the \textit{a posteriori} LLRs of
$\mathbf{x}^c_n$.

However, since the computational complexity of (\ref{eq:Lx2_n})
increases exponentially according to the length of feedback
filter, $L_d$, we seek a more practical modification. A possible
solution is to apply the Bayes' rule only for the two mutually
exclusive cases of $i_n = 0$ and $i_n \neq 0$. Then,
\begin{eqnarray}
\mathrm{Pr}(y_n  \mid x_n = + 1)  & = & \dfrac{\exp \left( L_e(x_n | i_n = 0 ) \right) \mathrm{Pr}(i_n = 0)}{1 + \exp \left(L_e(x_n | i_n =0 ) \right)} + \dfrac{\exp \left( L_e(x_n | i_n \neq 0 ) \right) \mathrm{Pr}(i_n \neq 0)}{1 + \exp \left( L_e(x_n | i_n \neq 0 ) \right) }   \\
\mathrm{Pr}(y_n  \mid x_n = - 1) & = & \dfrac{\mathrm{Pr}(i_n = 0)}{1 + \exp \left( L_e(x_n | i_n =0 ) \right)} + \dfrac{\mathrm{Pr}(i_n \neq 0)}{1 + \exp \left( L_e(x_n | i_n \neq 0 ) \right)}.
\end{eqnarray}
The extrinsic information of $x_n$ for each case of $i_n$ can be
estimated as
\begin{eqnarray}
L_e(x_n | i_n = 0 ) & = &  \dfrac{2 p_{\{n,0\}}  y_n   }{\mathrm{Var} (v_n) } \\
L_e(x_n | i_n \neq 0 ) & = &  \ln \left\{ \sum\limits_{j=1, \mathbf{e}^c_{\{n,j\}} \neq \mathbf{0}} ^{2^{L_d}} \dfrac{ \exp \left( L_e(x_n | \mathbf{e}^c_{\{n,j\}} ) \right) \mathrm{Pr}(\mathbf{e}^c_{\{n,j\}} ) } { \left\{ 1 + \exp \left( L_e(x_n | \mathbf{e}^c_{\{n,j\}} ) \right) \right\} \mathrm{Pr}(i_n \neq 0 ) } \right\} \nonumber \\
& &  - \ln \left\{ \sum\limits_{j=1, \mathbf{e}^c_{\{n,j\}} \neq \mathbf{0}}^{2^{L_d}} \dfrac{ \mathrm{Pr}(\mathbf{e}^c_{\{n,j\}}) } { \left\{ 1 + \exp \left( L_e(x_n | \mathbf{e}^c_{\{n,j\}} )  \right) \right\} \mathrm{Pr}(i_n \neq 0 ) } \right\} \nonumber \\
& \simeq & \ln \left\{ \sum\limits_{j=1, \mathbf{e}^c_{\{n,j\}} \neq \mathbf{0}} ^{2^{L_d}}  \left( \dfrac{1}{2} + \dfrac{ L_e(x_n | \mathbf{e}^c_{\{n,j\}} )}{4} \right ) \dfrac{ \mathrm{Pr}(\mathbf{e}^c_{\{n,j\}} ) } {  \mathrm{Pr}(i_n \neq 0 ) } \right\} \nonumber \\
& &  - \ln \left\{ \sum\limits_{j=1, \mathbf{e}^c_{\{n,j\}} \neq \mathbf{0}}^{2^{L_d}}  \left( \dfrac{1}{2} - \dfrac{ L_e(x_n | \mathbf{e}^c_{\{n,j\}} )}{4} \right ) \dfrac{ \mathrm{Pr}(\mathbf{e}^c_{\{n,j\}} ) } {  \mathrm{Pr}(i_n \neq 0 ) } \right\} \label{simeq:taylor1} \\
& = & \ln \left\{ \mathop\mathrm{E}\limits_{i_n} \left(  \dfrac{1}{2} + \dfrac{ 2 p_{\{n,0\}}  \left( y_n - i_n \right) }  {  4 \mathrm{Var} (v_n) } \Bigg | i_n \neq 0 \right) \right\}  - \ln \left\{\mathop\mathrm{E}\limits_{i_n} \left( \dfrac{1}{2} -\dfrac{2 p_{\{n,0\}}  \left( y_n - i_n \right) }  {  4 \mathrm{Var} (v_n) } \Bigg | i_n \neq 0 \right) \right\} \nonumber \\
& = & \ln \left\{ 1 + \dfrac{ p_{\{n,0\}}  \left( y_n - \mathrm{E}(i_n | i_n \neq 0 ) \right) }  {  \mathrm{Var} (v_n) } \right\}  - \ln \left\{ 1 - \dfrac{ p_{\{n,0\}}  \left( y_n - \mathrm{E}(i_n | i_n \neq 0 ) \right) }  {  \mathrm{Var} (v_n) } \right\} \nonumber \\
& \simeq & \left\{ {\begin{array}{*{20}c}   { 2\varphi_n / (1-\varphi_n) \textrm{ if $ \varphi_n < 0 $ } }  \\   { 2\varphi_n / (1+\varphi_n) \textrm{ otherwise} }  \\
\end{array}} \right. \label{simeq:L} \\
& = &  \dfrac{2\varphi_n}{1+|\varphi_n|}
\end{eqnarray}
where $\varphi_n \triangleq {p_{\{n,0\}}  \left( y_n - \mathrm{E}(i_n |
i_n \neq 0 ) \right)  } / { \mathrm{Var} (v_n) }$, $\mathrm{E}(i_n
| i_n \neq 0 ) = \mathrm{E}(i_n) / \mathrm{Pr}(i_n \neq 0 )$,
$\mathrm{Pr}(i_n = 0) = \prod_{k=1}^{L_d} \exp(| L(x_{n-k})|) /
(1+\exp(| L(x_{n-k})|) )$, and $\mathrm{Pr}(i_n \neq 0) = 1 -
\mathrm{Pr}(i_n = 0)$. The approximation of (\ref{simeq:taylor1})
is from the first order Taylor expansion at zero, i.e, $e^x /
(1+e^x) \simeq 0.5 + 0.25 x$ and $1 / (1+e^x) \simeq 0.5 - 0.25
x$. Furthermore, we also use $\ln \left\{ 1 + \varphi_n \right\}  - \ln
\left\{ 1 - \varphi_n \right\} =   \ln \left\{ 1+ 2\varphi_n / (1-\varphi_n) \right\} = -
\ln \left\{ 1 - 2\varphi_n / (1+\varphi_n) \right\}$ and $\ln(1+x) \simeq x$ in
(\ref{simeq:L}). In other words, $\ln \left\{ 1+ 2\varphi_n / (1-\varphi_n)
\right\} \simeq 2\varphi_n / (1-\varphi_n)$ is used for $\varphi_n<0$ while $-\ln \left\{ 1-
2\varphi_n / (1+\varphi_n) \right\} \simeq 2\varphi_n / (1+\varphi_n)$ is used for $\varphi_n \geq 0$.

Finally, the extrinsic information of $x_n$ is given as
\begin{eqnarray}
L_e(x_n) & = & \ln \left\{ \dfrac{ \exp \left( L_e(x_n | i_n = 0 ) \right) \mathrm{Pr}(i_n = 0) } { 1 + \exp \left( L_e(x_n | i_n =0 ) \right) } + \dfrac{ \exp \left( L_e(x_n | i_n \neq 0 ) \right) \mathrm{Pr}(i_n \neq 0) } { 1 + \exp \left( L_e(x_n | i_n \neq 0 ) \right) } \right \} \nonumber \\
& & - \ln \left\{ \dfrac{ \mathrm{Pr}(i_n = 0) } { 1 + \exp \left( L_e(x_n | i_n =0 ) \right) } + \dfrac{  \mathrm{Pr}(i_n \neq 0) } { 1 + \exp \left( L_e(x_n | i_n \neq 0 ) \right) } \right \}. \label{eq:Lx3_n}
\end{eqnarray}
While this gets passed to the outer decoder as equalizer's
extrinsic information, hard decisions that propagate down the
feedback filter are generated by slicing $L_e(x_n)+L_a(x_n)$ where
$L_a(x_n)$ is the extrinsic information from the decoder.

\subsection{Time-Invariant Filters}\label{TIV}
As also discussed in \cite{TE02}, the filter tap values derived
above are time-varying and creates significant implementation
challenges. A low-complexity variation would be to simply assume
the classical (non-turbo) DFE forward and feedback filter tap
solutions as in
\begin{eqnarray}
\mathbf{c}  & \triangleq &\left[ {c_{0} ,c_{+ 1} , \ldots , c_{L_c } } \right]^T  \nonumber  \\
 & = & \left( {\mathbf{H}\mathbf{\Sigma} \mathbf{H}^T + N_0 \mathbf{I} } \right)^{ - 1} \mathbf{s} \label{eq:tiv_c} \\
\mathbf{d}  & \triangleq & \left[ {d_{- L_d }, d_{- L_d  + 1} , \ldots , d_{- 1} } \right]^T  \nonumber \\
& = & \mathbf{M}\mathbf{H}^T\mathbf{c}, \label{eq:tiv_d}
\end{eqnarray}
where ${\mathbf{\Sigma}} \triangleq {\rm{Diag}} (\mathbf{0}_{1
\times L_d }, \mathbf{1}_{1 \times (L_c + 1) } )$, but let the
effect of decoder feedback come into play through the
subtraction of $\mathbf{H} \bar{\mathbf{x}}_n - \mathrm{E}(x_n)\mathbf{s}$ from the channel
observation vector (see (\ref{eq:y_n_2})) and the enhanced
\textit{a posteriori} LLR computation: $L_e(x_n)+L_a(x_n)$ where
$L_a(x_n)$ represents the decoder feedback.

By an obvious modification of (\ref{eq:new_model}), the equalized
signal is obtained as
\begin{eqnarray}
y_n & = & p_{0} x_n + i_n + v_n
\end{eqnarray}
where $p_{0} = \mathbf{c}^T \mathbf{s}$, $i_n = \sum_{k = 1}^{L_d}
d_{-k}  \big( { x_{n - k}  - \hat{x}_{n - k} } \big)$, $v_n =
\sum_{k = 1}^{L_c } p_{k} ( {x_{n + k}  - \mathrm{E}(x_{n + k} )}
) + \sum_{k = 0}^{L_c} c_{k} w_{n+k}$, and $\mathbf{p} \triangleq
\left[ {p_0 ,p_1 , \ldots ,p_{L_c} } \right] = \mathbf{c}^T
\mathbf{H}_1$. The mean and variance of $i_n$ and the noise
variance of $v_n$ with the time-invariant filters are also given
by
\begin{eqnarray}
\mathrm{E}(i_n) & = &\mathbf{d}^T \left( \tanh ( L(\mathbf{x}^c_n) /2) - \mathbf{\hat{x}}^c_n \right) \\
\mathrm{Var}(i_n) & = & \mathbf{d}^T \mathbf{\acute{\Sigma} }^c_n \mathbf{d} \\
\mathrm{Var}(v_n) & = &  \mathbf{c}^T \left(  \mathbf{H}\mathbf{\Sigma}_n \mathbf{H}^T   - z_n \mathbf{s}\mathbf{s}^T
+ N_0 \mathbf{I} \right) \mathbf{c} . \label{eq:time_invarinat_noise}
\end{eqnarray}

\section{SNR Advantage of BiDFE}\label{analysis1}
The idea of BiDFE is already motivated in \cite{BiDFE00},
\cite{BAD05} by the fact that DFE can be performed on the reversed
received sequence using the time-reversed channel response. Here
we derive the SNR figure-of-merit for BiDFE assuming ideal
feedback in both ways and allowing infinitely long filter lengths.
We then compare the result with those of the usual, single-sided
DFE as well as the matched filter detector (i.e., ideal detector
under zero-ISI condition). As will be seen, the ideal BiDFE SNR is
significantly better than the ideal DFE SNR especially at high
channel SNRs, further motivating a turbo BiDFE scheme.

\subsection{Unbiased MMSE-DFE}
It is well known that the $D$-transforms of the feedforward and
feedback MMSE-DFE filter coefficients are, respectively
\cite{Cioffi95}:
\begin{eqnarray}
c(D) & = & \dfrac{P_x}{P_0 g^*(D^{-*})}, \qquad d(D) = g(D)
\end{eqnarray}
where $P_0$ is such that $\log P_0 = \frac{1}{2\pi}
\int_{-\pi}^{\pi} \log R_{ss}(e^{-j\theta}) d\theta$ and
$g^*(D^{-*})$ is obtained from spectral factorization:
$R_{ss}(D) = P_x R_{hh}(D) + N_0 = P_0 g(D) g^*(D^{-*})$ where
$R_{hh}(D) = h(D)h^*(D^{-*})$ and $h(D)$ is the $D$-transform of the channel impulse response.

The unbiased equalized outputs of the normal MMSE-DFE in the
forward direction, $Y_f(D)$, are given by
\begin{eqnarray}
Y_f(D) & = & x(D) + \dfrac{P_0} {P_0 - N_0} e'_f(D) \label{eq:UDFE_out}
\end{eqnarray}
where \setlength\arraycolsep{1pt} \begin{eqnarray}
e'_f(D)  & \triangleq & \dfrac{N_0}{P_0}  \left ( 1 -
\dfrac{1}{g^{*}(D^{-*}) } \right ) x(D) + \dfrac{P_x w'(D)}{P_0
g^*(D^{-*})}  \label{eq:e_f}
\end{eqnarray}
with $w'(D)$ denoting a complex-valued Gaussian noise sequence with
autocorrelation function $R_{w'w'}(D) = N_0R_{hh}(D)$. Then, the mean-squared-error (MSE)
and SNR of the unbiased normal MMSE-DFE are given by
\setlength\arraycolsep{1pt}
\begin{eqnarray} \mathrm{MSE}_{UDFE} &
= & \left ( \dfrac {P_0} {P_0 - N_0} \right )^2
\mathrm{E}(|e'_{f,n}|^2)
 =  \dfrac{P_x N_0}{P_0 -N_0 }  \label{eq:MSE_UDFE}  \\
\mathrm{SNR}_{UDFE} & \triangleq & \dfrac{P_x}{\mathrm{MSE}_{UDFE}} = \dfrac{P_0 -N_0 }{N_0}.
\label{eq:SNR_UDFE}
\end{eqnarray}

\subsection{Unbiased Time-Reversed MMSE-DFE}
Now, let us assume that the transmitted data sequence $x_n$ is of
a finite length so that the MMSE-DFE can be performed on the
time-reversed received signals using the time-reverse of the
original channel impulse response \cite{RDFE92}. Denoting the
time-reversed ISI channel coefficients as $\tilde{h}_n =
h^*_{L_h-1-n}$, its \textit{D}-transform is given as $\tilde{h}(D)
= D^{L_h-1} h^{*}(D^{-*})$. Therefore, the $D$-transform of the
autocorrelation function of the time-reversed channel is given by
$R_{\tilde{h}\tilde{h}}(D) = \tilde{h}(D) \tilde{h}^*(D^{-*}) =
R_{hh}(D)$. Accordingly, the feedforward and feedback filters of
the time-reversed MMSE-DFE, denoted by $\tilde{c}(D)$ and
$\tilde{d}(D)-1$ respectively, are identical to the normal
MMSE-DFE filters, i.e.,
\begin{eqnarray}
\tilde{c}(D) & = & c(D) = \dfrac{P_x}{P_0 g^*(D^{-*})}, \qquad
\tilde{d}(D) = d(D) = g(D).
\end{eqnarray}

The unbiased output of the time-reversed MMSE-DFE can be expressed
similarly to the case of the normal, forward MMSE-DFE except that
the unbiased output sequence right after the time-reversed
MMSE-DFE should also be time-reversed, in order to get the
unbiased equalized output $Y_b(D)$ matched to the input sequence
$x(D)$. Therefore,
\begin{eqnarray}
Y_b(D) & = & x(D) + \dfrac{P_0} {P_0 - N_0}  e'_b(D) \label{eq:URDFE_out}
\end{eqnarray}
where
\begin{eqnarray}
e'_b(D)& \triangleq & \dfrac{N_0}{P_0}  \left ( 1 - \dfrac{1}{g(D) } \right ) x(D) + \dfrac{P_x}{P_0} \left ( \dfrac{w'(D)}{g(D)} \right ) \label{eq:e_b}.
\end{eqnarray}
Then, the MSE and SNR of the unbiased time-reversed MMSE-DFE are given by
\begin{eqnarray}
\mathrm{MSE}_{URDFE} & = & \left ( \dfrac{P_0}{P_0 - N_0} \right )^2 \mathrm{E}(|e'_{b,n}|^2) = \dfrac{P_x N_0}{P_0 -N_0 }   \label{eq:MSE_URDFE}  \\
\mathrm{SNR}_{URDFE} & \triangleq & \dfrac{P_x}{\mathrm{MSE}_{URDFE}} = \dfrac{P_0 - N_0 }{N_0}.
\label{eq:SNR_URDFE}
\end{eqnarray}

\subsection{Unbiased BiDFE}
\begin{figure}[!t] \centering
\includegraphics[width=11.0cm]{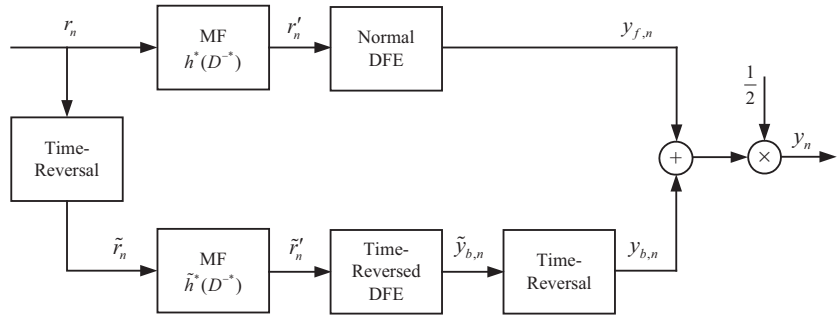}
\caption{Bidirectional Decision Feedback
Equalizer: Infinite Length.}\label{fig:BiDFE}
\end{figure}

The structure of the BiDFE is shown in Fig. \ref{fig:BiDFE}.
If we assume that the feedback sequence is correct, the outputs of
two unbiased DFEs are:
\begin{eqnarray}
 Y_{f,n} & = & X_n + V_{f,n}  \\
 Y_{b,n} & = & X_n + V_{b,n}
\end{eqnarray}
where $V_{f,n}$ and $V_{b,n}$ have \textit{D}-transforms
$V_f(D)$ and $V_b(D)$ as given by (from (\ref{eq:UDFE_out}), (\ref{eq:e_f}), (\ref{eq:URDFE_out}), and (\ref{eq:e_b}))
\begin{eqnarray}
V_f(D) %& \triangleq & \dfrac{P_0} {P_0 - N_0} e'_f(D) \nonumber \\
& = & \dfrac{N_0}{P_0 - N_0}  \left ( 1 - \dfrac{1}{g^{*}(D^{-*}) } \right ) x(D) + \dfrac{P_x}{P_0 - N_0} \left ( \dfrac{ w'(D)}{ g^*(D^{-*})} \right )  \label{eq:V_f(D)} \\
V_b(D) %& \triangleq & \dfrac{P_0} {P_0 - N_0} e'_b(D) \nonumber \\
& = & \dfrac{N_0}{P_0 - N_0}  \left ( 1 - \dfrac{1}{g(D) } \right ) x(D) + \dfrac{P_x}{P_0 - N_0} \left ( \dfrac{w'(D)}{g(D)} \right ) . \label{eq:V_b(D)}
\end{eqnarray}

Assuming stationary random processes, we drop time index $n$ for
notational simplicity and write: $Y_{f} = X + V_{f}$ and $ Y_{b} =
X + V_{b}$. From (\ref{eq:MSE_UDFE}) and (\ref{eq:MSE_URDFE}), the
variance of $V_{f}$ and $V_{b}$ are also given as:
\begin{eqnarray}
\mathrm{Var}(V_f) =  \mathrm{Var}(V_b) = \dfrac{P_x N_0}{P_0 - N_0}. \nonumber
\end{eqnarray}
The variables $V_{f}$ and $V_{b}$ are correlated with the
correlation coefficient given by
\begin{eqnarray}
\rho  & \triangleq & \dfrac{\mathrm{E} ( V_{f} V^{*}_{b} )
}{\sqrt{\mathrm{Var}(V_f) \mathrm{Var}(V_b) }} \nonumber \\
 & = & \dfrac{P_0 - N_0} {P_x N_0} \mathrm{E} \left [V_f(D) V^{*}_b(D^{-*}) \right ]_{0} \nonumber \\
& = & \dfrac{P_x} {N_0 (P_0 - N_0)} \mathrm{E}  \left [ \left ( \dfrac{1}{g^*(D^{-*})} \right )^2 w'(D) w'^*(D^{-*}) \right ]_{0}  \label{eq:co_x} \\
& = & \dfrac{P_0^2} { P_x N_0 (P_0 - N_0)} \left [  \left \{ c(D) \right \}^2  R_{w'w'}(D) \right]_{0} \nonumber \\
& = & \dfrac{P_0^2} {P_x(P_0 - N_0)} \left [\left \{ c(D) \right \}^2 R_{hh}(D) \right ]_{0} \label{eq:rho_inf}
\end{eqnarray}
where $\left [ z(D)\right ]_{0} = z_0 $ with $z(D) = \sum_k z_k
D^k$. The equality in (\ref{eq:co_x}) holds due to the assumption
that $X_n$ is an i.i.d random variable and the self-interference
term is removed from the expression $ 1 - 1/ g^{*}(D^{-*}) $.

Since $\mathrm{Var}(V_f) = \mathrm{Var}(V_b)$, the linear MMSE
combiner of \cite{BiDFE00}, \cite{Balakrishnan00} becomes $Y =
\frac{1}{2} \left( Y_f + Y_b \right)$. Naturally, the MSE and SNR
of the unbiased BiDFE are given as
\begin{eqnarray}
\mathrm{MSE}_{UBiDFE} & = & \dfrac{(1+\mathrm{Re}[\rho])}{2} \mathrm{MSE}_{UDFE}  =  \dfrac{(1+\mathrm{Re}[\rho]) P_x N_0}{2 (P_0 - N_0)}  \\
\mathrm{SNR}_{UBiDFE} & \triangleq & \dfrac{P_x}{\mathrm{MSE}_{UBiDFE}} =\dfrac{2}{(1+\mathrm{Re}[\rho])} \mathrm{SNR}_{UDFE}  =  \dfrac{2(P_0 -N_0) }{ (1 + \mathrm{Re}[\rho]) N_0}
\end{eqnarray}
where $\mathrm{Re}[\rho]$ denotes the real part of $\rho$.

Note that the infinite-length normal/time-reversed MMSE-DFE and
BiDFE analyzed here do not exploit the \textit{a priori}
information of $X_n$. In other words, the feedforward and feedback
filters of DFE are derived by assuming $\mathrm{E}(X_n) = 0$ for
all $n$, meaning that the calculated SNR performance would reflect
the non-turbo ideal-decision BiDFE performance with time-invariant
filter taps of Section \ref{TIV}.

\section{Derivation of Iterative BiDFE Algorithm}\label{BiDFEs}
\begin{figure}[!t]
\centering
\includegraphics[width=12.0cm]{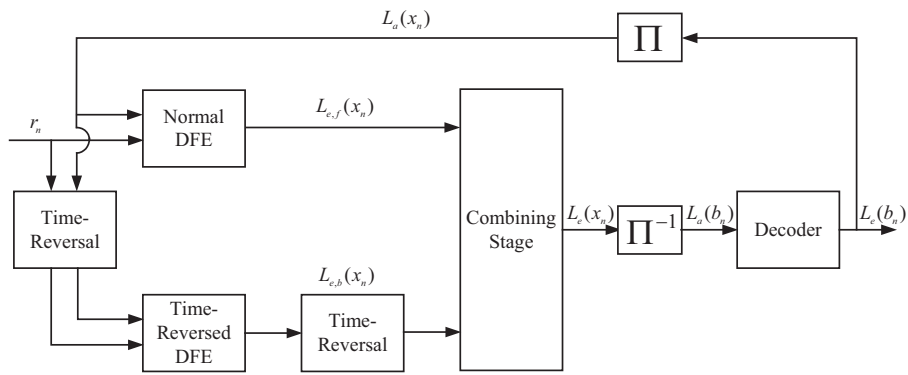}
\caption{Iterative Equalization Scheme based on BiDFE.}\label{fig:iterBiDFE}
\end{figure}

We now discuss an iterative BiDFE algorithm. Iterative
equalization schemes based on BiDFE are shown in Fig.
\ref{fig:iterBiDFE}. Basically, the channel equalizer is a SISO
equalizer which employs the normal forward DFE, the time-reversed
DFE and an LLR combining block. The received data sequence is
equalized in both directions by the two DFEs, and the extrinsic
information from two DFEs are combined and passed to the error
correction code decoder. We show that a proper combining of the
two sets of extrinsic information can suppress error propagation
and noise further and generate more reliable extrinsic information
for the outer decoder.

\subsection{Combining Extrinsic Information}
Similarly to the finite-length time-varying feedforward and
feedback filter of the normal DFE at time index $n$, which are
previously defined as $\mathbf{c}_n$ in (\ref{eq:c_n}) and
$\mathbf{d}_n$ in (\ref{eq:d_n}), we also define the finite-length
time-varying feedforward and feedback filter of the time-reversed
DFE at time index $n$ as $\mathbf{\tilde{c}}_n$ and
$\mathbf{\tilde{d}}_n$ with the same lengths as $\mathbf{c}_n$ and
$\mathbf{d}_n$ respectively. Note that $\mathbf{\tilde{c}}_n$ and
$\mathbf{\tilde{d}}_n$ are defined in a similar way as
(\ref{eq:c_n}) and (\ref{eq:d_n}) except that the channel
convolution matrix $\mathbf{\tilde{H}}$ for the time-reversed
channel is given as
\begin{equation} \mathbf{\tilde{H}} \triangleq \left [ {\begin{array}{*{20}c}
   {h_0 } & {h_{1} } &  \cdots  & {h_{L_h  - 1} } & 0 &  \cdots  & {} & 0  \\
   0 & {h_0 } & {h_1 } &  \cdots  & {h_{L_h  - 1} } & 0 &  \cdots  & 0  \\
   {} & \ddots & {} &  \ddots  & {} & \ddots & {} & {}  \\
   0 & 0 &  \cdots  & 0 & {h_0 } & {h_1 } &  \cdots  & {h_{L_h  - 1} }  \\
\end{array}} \right ]. \nonumber
\end{equation}

The unbiased equalizer output
\cite{Cioffi95} corresponding to the transmitted coded symbol from
the the normal (forward) and the time-reversed (backward) DFE can
be represented respectively as
\begin{eqnarray}
Y_{f,n} & = & X_n + I_{f,n} + V_{f,n} \label{eq:U_{f,n}} \\
Y_{b,n} & = & X_n + I_{b,n} + V_{b,n} \label{eq:U_{b,n}}
\end{eqnarray}
where $X_n \triangleq x_n$, $V_{f,n} \triangleq
{v_{f,n}}/{p_{\{n,0\}}}$ and $ I_{f,n} \triangleq
{i_{f,n}}/{p_{\{n,0\}}}$. Also, $V_{b,n} \triangleq
{v_{b,n}}/{\tilde{p}_{\{n,0\}}}$ and $ I_{b,n} \triangleq
{i_{b,n}}/{\tilde{p}_{\{n,0\}}}$ where $v_{b,n}$ and $i_{b,n}$ are
defined similarly to the normal DFE and $\tilde{p}_{\{n,0\}} =
\mathbf{\tilde{c}}_n^T \mathbf{\tilde{s}} $ where
${\mathbf{\tilde{s} }} \triangleq {\mathbf{\tilde{H} }}
[\mathbf{0}_{1 \times L_d } , 1 , \mathbf{0}_{1 \times L_c } ]^T$.
For notational simplicity, we further drop time index $n$ with an
understanding that processing remains identical as $n$ progresses:
$Y_{f} = X + I_{f} + V_f$ and $Y_{b} =  X + I_{b} + V_b$.

Now, we discuss the problem of how to combine the extrinsic
information from two DFEs. Initially, let us consider two unbiased
equalizer outputs, which are corrupted by AWGN, corresponding to
the transmitted coded symbol $X$:
\begin{eqnarray}
Y_{f} & = & X + U_{f} \nonumber \\
Y_{b} & = & X + U_{b} \nonumber
\end{eqnarray}
where the noise $U_f$ and $U_b$ are assumed to be zero mean
Gaussian random variables which are independent of the coded
data $X$ but correlated with each other with correlation
coefficient $\rho$.

In order to combine the extrinsic information, it is beneficial to
whiten the noise $U_f$ and $U_b$ before combining. The noise
correlation matrix $\mathbf{R}$ is defined as
\begin{eqnarray}
\mathbf{R} & \triangleq & \left[ {\begin{array}{*{20}c}
   {\mathrm{Var}(U_f) } & {\mathrm{E}(U_f U_b) }  \\
   {\mathrm{E}(U_f U_b) } & {\mathrm{Var}(U_b) }  \\
\end{array}} \right]  =  \left[ {\begin{array}{*{20}c}
   {N_f } & {\rho \sqrt{N_f N_b} }  \\
   {\rho \sqrt{N_f N_b} } & {N_b }  \\
\end{array}} \right] \nonumber
\end{eqnarray}
where $N_f \triangleq \mathrm{Var}(U_f)$ and $N_b \triangleq
\mathrm{Var}(U_b)$. Then, the eigenvalues of the noise correlation
matrix, $\lambda_{1}$ and $\lambda_{2}$, with their corresponding
normalized eigenvectors $\mathbf{g}_{1}$ and $\mathbf{g}_{2}$ are
given by
\begin{eqnarray}
& & \lambda_{1} = \dfrac{ (N_f + N_b) + \sqrt{ (N_f-N_b)^2 + 4\rho^2 N_f N_b }}{2} \nonumber \\
& & \lambda_{2} = \dfrac{ (N_f + N_b) - \sqrt{ (N_f-N_b)^2 + 4\rho^2 N_f N_b }}{2} \nonumber \\
& & \mathbf{g}_1 = \dfrac{1}{\sqrt{g_{11}^2 + g_{21}^2}} \left[  {\begin{array}{*{20}c}
   {g_{11}}  \\
   {g_{21}}  \\
\end{array}} \right], \:
\mathbf{g}_2 = \dfrac{1}{\sqrt{g_{12}^2 + g_{22}^2}} \left[  {\begin{array}{*{20}c}
   {g_{12}}  \\
   {g_{22}}  \\
\end{array}} \right]  \nonumber
\end{eqnarray}
where $g_{11} =  \frac{1}{2} \Big[ (N_f - N_b) + \sqrt{
(N_f-N_b)^2 + 4\rho^2 N_f N_b } \Big]$, $g_{12} =  \frac{1}{2} \Big[ (N_f - N_b) - \sqrt{
(N_f-N_b)^2 + 4\rho^2 N_f N_b } \Big]$, and $g_{21} = g_{22} =
\rho \sqrt{N_f N_b}$. It is easy to see that the noise correlation matrix
$\mathbf{R}$ is non-singular unless $\rho = \pm 1$. If
$\mathbf{R}$ is non-singular, $\mathbf{R}$ can be expanded as
$\mathbf{R} = \mathbf{G} \mathbf{\Lambda} \mathbf{G}^{-1}$ where $
\mathbf{G} \triangleq  \left[ \mathbf{g}_{1} \: \mathbf{g}_{2}
\right]$ and $\mathbf{\Lambda}  \triangleq
\mathrm{Diag}(\lambda_1, \lambda_2)$. Since $\mathbf{G}$ is a
unitary matrix, the noise whitening matrix is $\mathbf{A}
\triangleq \left[ \mathbf{a}_{1} \: \mathbf{a}_{2} \right] =
\mathbf{G}^{-1} = \mathbf{G}^{T}$ where $\mathbf{a}_{1} \triangleq
\left[ a_{11} \: a_{21} \right]^T$ and $\mathbf{a}_{2} \triangleq
\left[ a_{12} \: a_{22} \right]^T$. So, given the equalized output
vector $\mathbf{Y} \triangleq [Y_f, Y_b]^T$, the whitened vector
is $\mathbf{Y}' \triangleq  [Y'_f, Y'_b]^T  =
\mathbf{A}\mathbf{Y}$ with the new noise correlation matrix
$\mathbf{R}' = \mathbf{A} \mathbf{R} \mathbf{A}^{T} =
\mathbf{\Lambda}$. Finally, the extrinsic information of $X$ can
be expressed as
\begin{eqnarray}
L_e(X) %& \triangleq &  \ln \dfrac{\mathrm{Pr}( X = +1 \mid Y_f , Y_b ) }{\mathrm{Pr}(X = -1 \mid Y_f , Y_b)}  \nonumber \\
& = & \ln \dfrac{\mathrm{Pr}(Y_f, Y_b  \mid X = +1)}{\mathrm{Pr}(Y_f, Y_b \mid X = -1)} \nonumber \\
& = & \ln \dfrac{\mathrm{Pr}(Y'_f, Y'_b  \mid X = +1)}{\mathrm{Pr}(Y'_f, Y'_b \mid X = -1)} \nonumber \\
& = & \ln \dfrac{\mathrm{Pr}(Y'_f \mid X = +1)}{\mathrm{Pr}(Y'_f \mid X = -1)} + \ln \dfrac{\mathrm{Pr}(Y'_b \mid X = +1)}{\mathrm{Pr}(Y'_b \mid X = -1)}\nonumber \\
& = & \dfrac{ 2(a_{11}+a_{12}) Y'_f } {\lambda_1} + \dfrac{ 2(a_{21}+a_{22}) Y'_b } {\lambda_2} \nonumber \\
%& = & \dfrac{ 2(a_{11}+a_{12}) (a_{11}Y_f  + a_{12} Y_b ) } {\lambda_1} + \dfrac{ 2(a_{21}+a_{22}) (a_{21}Y_f  + a_{22} Y_b) } {\lambda_2} \nonumber \\
& = & \dfrac{ 2\left( N_b - \rho \sqrt{N_f N_b} \right)Y_f  } { \left(1-\rho^2 \right) N_f N_b } + \dfrac{ 2\left( N_f - \rho \sqrt{N_f N_b} \right) Y_b } { \left(1-\rho^2 \right) N_f N_b }   \nonumber \\
& = & \dfrac{ \left( N_b - \rho \sqrt{N_f N_b} \right) } { \left(1-\rho^2 \right) N_b } L_{e,f}(X) + \dfrac{ \left( N_f - \rho \sqrt{N_f N_b} \right) } { \left(1-\rho^2 \right) N_f } L_{e,b}(X) .
\label{eq:combiner3}
\end{eqnarray}

For the singular noise correlation matrix $\mathbf{R}$ (i.e.,
$\rho = +1$), $N_f = N_b = N$ and $Y_f = Y_b = Y$ so that
$L_{e,f}(X) = L_{e,b}(X)$. Consequently, the extrinsic information
of $X$ becomes $L_e(X) = 2 Y /N = (L_{e,f}(X) + L_{e,b}(X)) / 2$.
Note that the mean combiner of \cite{BiSFE06}, $L_e(X) =
(L_{e,f}(X) + L_{e,b}(X))/2$, can be considered as the proposed
combiner with $\rho = +1$. If $\rho = -1$, $U_f = -U_b$ and we can
cancel out the noise perfectly by averaging the outputs: $(Y_f +
Y_b) /2$. The extrinsic information of $X$ in this case is
$L_e(X)= +\infty$ when $(Y_f + Y_b) /2 \geq 0 $ while $L_e(X)=
-\infty$ when $(Y_f + Y_b) /2 < 0$.

\subsection{Reducing the Combiner Sensitivity to the Estimation Error}
Let us consider the effect of errors in estimating $\rho$ on
extrinsic information. Write $\hat{\rho} = \rho + \varepsilon$
where $\varepsilon$ is the estimation error. Then, the sensitivity
of the combiner in (\ref{eq:combiner3}) to the estimation error
can be defined as
\begin{eqnarray}
\mathrm{S}(\rho) & \triangleq & \left |  \dfrac{\partial  L_e(X)} {\partial \rho}  \right | \nonumber \\
& = & \bigg | \dfrac{ \left( 2\rho N_b - (1 + \rho^2 ) \sqrt{N_f N_b} \right) } { \left(1-\rho^2 \right)^2  N_b } L_{e,f}(X)  + \dfrac{ \left( 2\rho N_f - (1 + \rho^2 ) \sqrt{N_f N_b} \right) } { \left(1-\rho^2 \right)^2  N_f } L_{e,b}(X) \bigg | \nonumber
\end{eqnarray}
which approaches infinity as $\rho \rightarrow \pm 1$. This means
that the combiner of (\ref{eq:combiner3}) is unfortunately very
sensitive to the correlation estimator error, as the magnitude of
the correlation becomes large.

The sensitivity of the combiner can be reduced if we assume that
the variance of $U_f$ and $U_b$ are the same, i.e., $N = N_f = N_b
= (N_f + N_b)/2$. This assumption is reasonable when the same
feedforward and feedback filter length is used in both DFEs. Then,
from (\ref{eq:combiner3}), the combined extrinsic information of
$X$ for non-singular $\mathbf{R}$ is simply given as
\begin{eqnarray}
L_e(X) =  \dfrac{1} {(1+ \rho)} \Big( L_{e,f}(X) + L_{e,b}(X) \Big) \label{eq:combiner5}
\end{eqnarray}
with the sensitivity to the correlation estimation error
\begin{eqnarray}
\mathrm{S}(\rho) =  \left | \dfrac{1}{(1+\rho)^2} \Big( L_{e,f}(X) + L_{e,b}(X) \Big) \right | . \nonumber
\end{eqnarray}
Although the sensitivity of this combiner to the estimation error
also goes to infinity as $\rho \rightarrow -1$, it shows more
robustness as $\rho \rightarrow +1$
since $ \lim_{\rho \rightarrow + 1} \mathrm{S}(\rho) =  \left |( L_{e,f}(X) + L_{e,b}(X)) /4 \right | $.

\subsection{Application to the BiDFE Algorithm}
In this paper, although the composite noise $I_{f,n} + V_{f,n}$ and $I_{b,n} + V_{b,n}$ are not Gaussian, we exploit the combiner of (\ref{eq:combiner5}) in order to produce the combined extrinsic information to be passed to the convolutional decoder. The noise correlation coefficient between $I_{f,n} + V_{f,n}$ and $I_{b,n} + V_{b,n}$ is naturally defined as
\begin{eqnarray}
\rho_n  & \triangleq & \dfrac{\mathrm{E} \left\{ \left(I_{f,n} - \mathrm{E}(I_{f,n}) + V_{f,n} \right) \left( I_{b,n} - \mathrm{E}(I_{b,n}) + V_{b,n} \right) \right\} }{\sqrt{ \left( \mathrm{Var}(I_{f,n}) + \mathrm{Var}(V_{f,n}) \right) \left( \mathrm{Var}(I_{b,n}) + \mathrm{Var}(V_{b,n}) \right) }}.
\end{eqnarray}

Unfortunately, it is difficult to compute the correlation
coefficient analytically in the presence of decision feedback
errors. However, assuming that the noise is stationary, we have
$\rho_n = \rho$ and the correlation coefficient can be estimated
through time-averaging:
\begin{eqnarray}
\hat{\rho} & = & \dfrac{\sum \left \{  (Y_{f,n} - \hat{X}_{f,n} - \mathrm{E}(I_{f,n}) ) ( Y_{b,n} - \hat{X}_{b,n} - \mathrm{E}(I_{b,n}) ) \right \} }{\sqrt{\sum (Y_{f,n} - \hat{X}_{f,n} - \mathrm{E}(I_{f,n}) )^2} \sqrt{\sum ( Y_{b,n} - \hat{X}_{b,n} - \mathrm{E}(I_{b,n}) )^2}} \label{eq:rho}
\end{eqnarray}
where the summations are over some reasonably large finite window.
Note that the hard decisions for the transmitted symbols in normal
and time-reversed DFEs might be different; in estimating the
correlation coefficient, we only consider those noise samples for
which $\hat{X}_{f,n}$ and $\hat{X}_{b,n}$ are identical.

Let us summarize our LLR combining method: 1) The extrinsic
information $L_{e,f}(X_n)$ and $L_{e,b}(X_n)$ for $n=1,2,\ldots,L$
are acquired according to (\ref{eq:Lx3_n}) in the normal and
time-reversed MMSE-DFE settings. 2) Estimate the noise correlation
coefficient, $\hat{\rho}$, between $I_{f,n} + V_{f,n}$ and
$I_{b,n} + V_{b,n}$ by (\ref{eq:rho}). 3) Generate the combined
extrinsic information $L_e(X_n)$ according to (\ref{eq:combiner5})
with $\rho_n=\hat{\rho}$.

\subsection{Correlation Analysis under Ideal Feedback}
We provide correlation analysis in the following. The analysis
will allow validation of (\ref{eq:rho}) in different scenarios.
The observation of how the simulated correlation coefficient
(\ref{eq:rho}) converges to the analytically computed one under
the assumptions of ideal feedback and perfect \textit{a priori}
information will also provide useful insights into the iterative
behaviour of the proposed turbo BiDFE.

First of all, the noise variance of $V_{f,n}$ and $V_{b,n}$ from the time-varying filters are:
\begin{eqnarray}
\mathrm{Var}(V_{f,n}) & = &  (1- \mathbf{s}^T \mathbf{c}_n) / \mathbf{c}_n^T \mathbf{s} \nonumber \\
\mathrm{Var}(V_{b,n}) & = &  (1- \mathbf{\tilde{s}}^T \mathbf{\tilde{c}}_n) / \mathbf{\tilde{c}}_n^T \mathbf{\tilde{s}}. \nonumber
\end{eqnarray}
When we assume ideal decision feedback, $\mathrm{Pr}(I_f = 0) =
\mathrm{Pr}(I_b = 0) = 1$ so that $I_{f,n} = I_{b,n} = 0$, the
noise correlation coefficient $\rho_n$ between $V_{f,n}$ and
$V_{b,n}$ becomes
\begin{eqnarray}
\rho_n  & \triangleq &  \dfrac{ \mathrm{E} ( V_{f,n} V_{b,n} ) }{\sqrt{\mathrm{Var}(V_{f,n}) \mathrm{Var}(V_{b,n}) } } \nonumber \\
& = & \dfrac{ \mathrm{E} \left [ \left\{ \dfrac{1}{p_{\{n,0\}}} \sum\limits_{j = 0}^{L_c} c_{\{n,j\}} w_{n+j} \right\} \left\{  \dfrac{1}{\tilde{p}_{\{n,0\}}} \sum\limits_{k = 0}^{L_c} \tilde{c}_{\{n,k\}} w_{n-k+L_h-1} \right\} \right ]} { \sqrt{(1- \mathbf{s}^T \mathbf{c}_n) / \mathbf{c}_n^T \mathbf{s}} \sqrt{(1- \mathbf{\tilde{s}}^T \mathbf{\tilde{c}}_n) / \mathbf{\tilde{c}}_n^T \mathbf{\tilde{s}} } } \label{eq:co_x2} \\
& = & \dfrac{  \sum\limits_{j = 0}^{L_c} \sum\limits_{k = 0}^{L_c}  c_{\{n,j\}} \tilde{c}_{\{n,k\}} \mathrm{E} \left [  w_{n+j} w_{n-k+L_h-1} \right ]} { \sqrt{ \mathbf{c}_n^T \mathbf{s} (1- \mathbf{s}^T \mathbf{c}_n) } \sqrt{ \mathbf{\tilde{c}}_n^T \mathbf{\tilde{s}}(1- \mathbf{\tilde{s}}^T \mathbf{\tilde{c}}_n) }  }  \nonumber \\
& = & N_0  \left( \dfrac{ \sum\limits_{j = 0}^{L_c} \sum\limits_{k = 0}^{L_c}  c_{\{n,j\}} \tilde{c}_{\{n,k\}} \delta(j+k+1-L_h) } { \sqrt{ \mathbf{c}_n^T \mathbf{s} (1- \mathbf{s}^T \mathbf{c}_n) } \sqrt{ \mathbf{\tilde{c}}_n^T \mathbf{\tilde{s}}(1- \mathbf{\tilde{s}}^T \mathbf{\tilde{c}}_n) }  }  \right) \label{eq:rho_tv}
\end{eqnarray}
where $\delta(t)$ is defined as: if $t=0$, $\delta(t)=1$;
otherwise, $\delta(t)=0$. The equality in (\ref{eq:co_x2}) holds
because $X_n$ is an i.i.d random variable.

If the time-invariant filters are used instead of the time-varying
filters, the variances of $V_{f,n}$ and $V_{b,n}$ become
\begin{eqnarray}
\mathrm{Var}(V_{f,n}) & = & \mathbf{c}^T \left(  \mathbf{H}\mathbf{\Sigma}_n \mathbf{H}^T  - z_n \mathbf{s}\mathbf{s}^T +{N_0 \mathbf{I}} \right) \mathbf{c} / \left( \mathbf{c}^T \mathbf{s} \right)^2 \nonumber \\
\mathrm{Var}(V_{b,n}) & = & \mathbf{\tilde{c}}^T \left(  \mathbf{\tilde{H}}\mathbf{\tilde{\Sigma}}_n \mathbf{\tilde{H}}^T  - z_n \mathbf{\tilde{s}}\mathbf{\tilde{s}}^T +{N_0 \mathbf{I}} \right) \mathbf{\tilde{c}} / \left( \mathbf{\tilde{c}}^T \mathbf{\tilde{s}} \right)^2. \nonumber
\end{eqnarray}
Then, the noise correlation coefficient can be also obtained as
\begin{eqnarray}
\rho_n  & = & N_0 \left( \dfrac{ \sum\limits_{j = 0}^{L_c} \sum\limits_{k = 0}^{L_c}  c_{j} \tilde{c}_{k} \delta(j+k+1-L_h) } { \sqrt{ \mathbf{c}^T \left( \mathbf{H}\mathbf{\Sigma}_n \mathbf{H}^T  - z_n \mathbf{s}\mathbf{s}^T +{N_0 \mathbf{I}} \right) \mathbf{c}} \sqrt{ \mathbf{\tilde{c}}^T (  \mathbf{\tilde{H}}\mathbf{\tilde{\Sigma}}_n \mathbf{\tilde{H}}^T  - z_n \mathbf{\tilde{s}}\mathbf{\tilde{s}}^T +{N_0 \mathbf{I}} ) \mathbf{\tilde{c}} }  } \right) . \label{eq:rho_tiv}
\end{eqnarray}

Now, let us consider some special cases.

\subsubsection{No \textit{A Priori} Information}
When no \textit{a priori} information is available, i.e.,
$\mathrm{E}(X_n) = 0$ for all $n$, the feedforward and feedback
filters are the same as the time-invariant filters and the noise
variances are stationary:
\begin{eqnarray}
\mathrm{Var}(V_{f,n}) & = & \mathrm{Var}(V_{f}) =(1- \mathbf{s}^T \mathbf{c}) / \mathbf{c}^T \mathbf{s} \nonumber \\
\mathrm{Var}(V_{b,n}) & = & \mathrm{Var}(V_{b}) = (1- \mathbf{\tilde{s}}^T \mathbf{\tilde{c}}) / \mathbf{\tilde{c}}^T \mathbf{\tilde{s}}. \nonumber
\end{eqnarray}
Therefore, the noise correlation coefficient is given by
\begin{eqnarray}
\rho_n  & = & \rho  = N_0 \left( \dfrac{\sum\limits_{j = 0}^{L_c} \sum\limits_{k = 0}^{L_c}  c_{j} \tilde{c}_{k} \delta(j+k+1-L_h) } { \sqrt{\mathbf{c}^T \mathbf{s} (1- \mathbf{s}^T \mathbf{c}) } \sqrt{\mathbf{\tilde{c}}^T \mathbf{\tilde{s}} (1- \mathbf{\tilde{s}}^T \mathbf{\tilde{c}}) } } \right). \label{eq:rho_fi}
\end{eqnarray}
We observed that the noise correlation coefficient of
the infinite-length BiDFE in (\ref{eq:rho_inf}) is almost
identical to that of the finite-length BiDFE in (\ref{eq:rho_fi})
when $L_c$ is chosen to be long enough.

\subsubsection{Time-varying Filters with Perfect \textit{A Priori} Information}
When several iterations are performed at high SNRs in turbo
equalization, the perfect \textit{a priori} information could be
available, i.e., $\mathrm{E}(X_n) = X_n$ for all $n$. When
$\mathrm{E}(X_n) = X_n$ for all $n$, the feedforward filters
$\mathbf{c}_n$ and $\mathbf{\tilde{c}}_n$ of two DFEs become the
normalized matched filters corresponding to the forward and
reverse channel impulse responses:
\begin{eqnarray}
\mathbf{c}_n & = & A \left[h_0, h_1, \ldots , h_{L_h-1}, \mathbf{0}_{1 \times L_c - L_h +1} \right]^T \nonumber \\
\mathbf{\tilde{c}}_n & = & A \left[h_{L_h-1}, h_{L_h-2}, \ldots , h_{0}, \mathbf{0}_{1 \times L_c - L_h +1} \right]^T \nonumber
\end{eqnarray}
where $A$ is a real-valued constant depending on SNR, i.e., $A
\triangleq 1/(N_0 + \sum_{k=0}^{L_h-1} |h_k|^2)$. Moreover, since
the first terms of $V_{f,n}$ and $V_{b,n}$ disappear, the noise
variances are simply:
\begin{eqnarray}
\mathrm{Var}(V_{f,n}) & = & \mathrm{Var}(V_{f}) =  N_0  \mathbf{c}_n^T  \mathbf{c}_n / (\mathbf{c}_n^T \mathbf{s})^2 = \dfrac{ N_0 A^2 } { (\mathbf{c}_n^T \mathbf{s})^2 } \sum_{k=0}^{L_h-1} |h_k|^2 \nonumber \\
\mathrm{Var}(V_{b,n}) & = & \mathrm{Var}(V_{b}) = N_0  \mathbf{\tilde{c}}_n^T  \mathbf{\tilde{c}}_n / (\mathbf{\tilde{c}}_n^T \mathbf{\tilde{s}})^2 = \dfrac{N_0 A^2 }{(\mathbf{\tilde{c}}_n^T \mathbf{\tilde{s}})^2 } \sum_{k=0}^{L_h-1} |h_k|^2 . \nonumber
\end{eqnarray}
Accordingly, the noise correlation coefficient is
\begin{eqnarray}
\rho_n  = \rho =  1. \label{eq:tv_rho2}
\end{eqnarray}
Note that the noise correlation coefficient $\rho$ with perfect
\textit{a priori} information converges to 1 regardless of the SNR
value. As will be shown shortly, the measured correlation
coefficient using simulated turbo BiDFE outputs indeed approaches
1, as turbo iteration progresses. This indicates that both
assumptions - ideal decision feedback and perfect \textit{a
priori} information - are reasonable.

\subsubsection{Time-invariant Filters with Perfect \textit{A Priori} Information}
When the time-invariant filters are used with perfect \textit{a
priori} information, the time-invariant DFEs yield the noise
variances as
\begin{eqnarray}
\mathrm{Var}(V_{f,n}) & = & \mathrm{Var}(V_{f}) =  N_0  \mathbf{c}^T  \mathbf{c} / (\mathbf{c}^T \mathbf{s})^2 \nonumber \\
\mathrm{Var}(V_{b,n}) & = & \mathrm{Var}(V_{b}) =  N_0  \mathbf{\tilde{c}}^T  \mathbf{\tilde{c}} / (\mathbf{\tilde{c}}^T \mathbf{\tilde{s}})^2. \nonumber
\end{eqnarray}
The noise correlation coefficient is also simply given by
\begin{eqnarray}
\rho_n  = \rho =  \dfrac{  \sum\limits_{j = 0}^{L_c} \sum\limits_{k = 0}^{L_c}  c_{j} \tilde{c}_{k} \delta(j+k+1-L_h) } { \sqrt{\mathbf{c}^T \mathbf{c}} \sqrt{\mathbf{\tilde{c}}^T \mathbf{\tilde{c}}} }  . \label{eq:tiv_rho2}
\end{eqnarray}
As will be discussed in the next section, in the simulation of
turbo BiDFE with time-invariant taps it is observed that the BiDFE
output correlation does indeed converge to (\ref{eq:tiv_rho2}),
indicating again that the assumptions of error-free decisions and
perfect \textit{a priori} information are reasonable.

\section{Simulation Results}\label{simulation results}
\begin{figure}[!t]
\centering % centers everything on the page
\includegraphics[width=15cm]{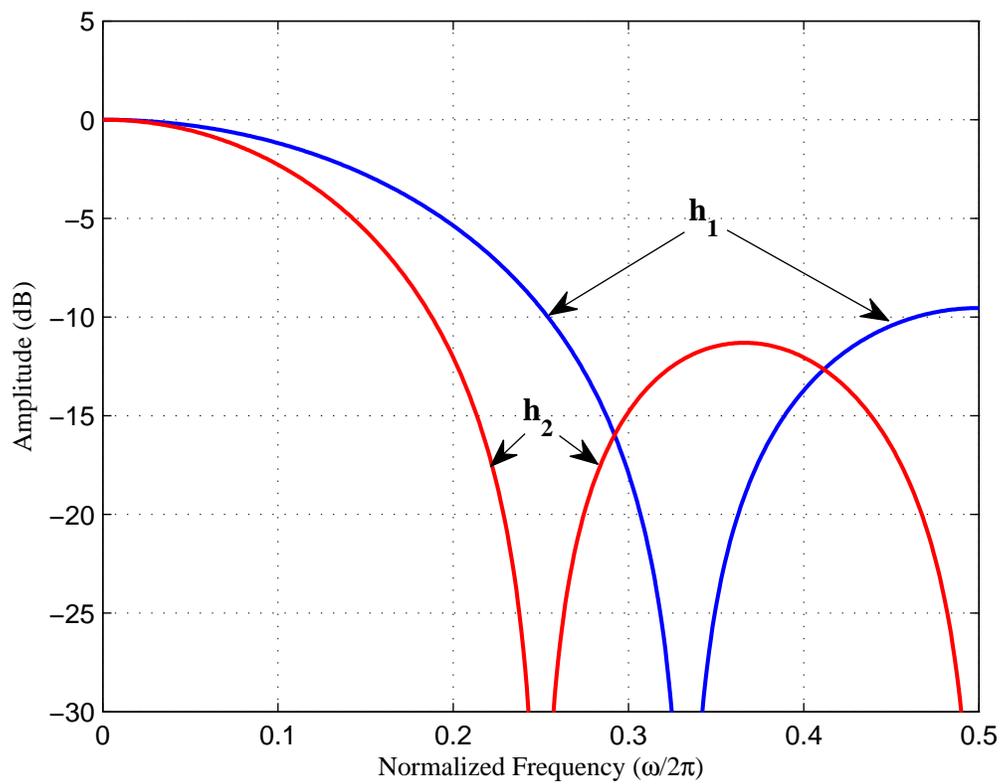}
\caption{Frequency Magnitude Response of the ISI Channels:
$\mathbf{h_1} =(1/\sqrt{19})
[1 \quad 2 \quad 3 \quad 2 \quad 1]^T$, $\mathbf{h_2}
=(1/\sqrt{44}) [1 \quad 2 \quad 3 \quad 4 \quad 3 \quad 2 \quad
1]^T$.}
 \label{fig:Freq_channels}
\end{figure}

In this section, simulation results of several iterative
equalization schemes are presented. The transmitted symbols are
encoded with a recursive rate-$1/2$ convolutional code encoder
with parity generator $(1+D^2) / (1+D+D^2)$ with $2^{11}$ message
bits and are modulated by binary phase-shift keying (BPSK) so that
$x_n \in \{\pm 1\}$. We also assume that the noise is AWGN, and
the noise variance and the channel information are perfectly known
to the receiver. The ISI channels with impulse responses
$\mathbf{h_1} =(1/\sqrt{19}) [1 \quad 2 \quad 3 \quad 2 \quad
1]^T$ and $\mathbf{h_2} =(1/\sqrt{44}) [1 \quad 2 \quad 3 \quad 4
\quad 3 \quad 2 \quad 1]^T$ investigated in \cite{TE02} and
\cite{Jae05} are used for evaluating the performance of the
iterative equalizers. These channels are considered very severe
ISI channels as the channel spectra possess nulls over the Nyquist
band, as shown in Fig. \ref{fig:Freq_channels}. Finally, the
decoder is implemented using the BCJR algorithm. Only the SISO
equalizer changes from one scheme to another. The MMSE-DFE with 17
feedforward taps and 4 feedback taps is used for both the normal
and the time-reversed DFEs on $\mathbf{h_1}$ while MMSE-DFE with
21 feedforward taps and 6 feedback taps is used on $\mathbf{h_2}$.
Finally, the linear MMSE equalizer uses 21 taps for $\mathbf{h_1}$
and 27 taps for $\mathbf{h_2}$.

Six different equalizer types are simulated in this work. The
notation ``TV-" denotes equalizers with time-varying filters while
``TIV-" indicates those with time-invariant filters. For instance,
``TV-LE" in the legend indicates the linear MMSE equalizer
with a time-varying filter. The ``Proposed DFE" uses the proposed
LLR mapping of (\ref{eq:Lx3_n}) while ``DFE" uses the conventional
LLR mapping (as used in \cite{TE02}) The ``Proposed BiDFE" is the
iterative BiDFE algorithm which is described in Section
\ref{BiDFEs}. In other words, ``Proposed BiDFE" uses the the
proposed LLR generation for both normal and time-reversed DFEs
along with the proposed extrinsic information combiner of
(\ref{eq:combiner5}) in conjunction with the noise correlation
coefficient of (\ref{eq:rho}). The ``BiDFE (mean combiner)" is the
iterative BiDFE algorithm with the conventional LLR mapping and
the mean combiner, $L_e(X) = (L_{e,f}(X) + L_{e,b}(X))/2$ (of
\cite{BiSFE06}), simulated for performance comparison purposes.
Finally, ``MAP" is the optimal equalizer implemented via the
BCJR algorithm.

A thorough comparison is given in \cite{TE02} on the required
complexity levels of the SISO-LE, SISO-DFE and the MAP equalizers.
The exact level of implementation complexity is hard to assess as
it depends highly on specific VLSI architecture details. Roughly
speaking, however, it is safe to say that the number of
multiplications and additions increases as an exponential function
of the channel memory length for the MAP equalizer whereas the
number of the same operations is a quadratic function of both the
channel memory length and the filter length for the TV-LE and the
TV-DFE, as shown in \cite{TE02}. The number of operations, on the
other hand, increases only linearly for the TIV-LE and the
TIV-DFE~\cite{TE02}. The BiDFE equalizers, including the proposed
BiDFE methods, require roughly twice as many operations as the DFE
counterparts, due to the presence of the time-reversed filter
components. Most notably, while the complexity of the proposed
BiDFE with time-invariant filters is considerably lower than that
of the MAP equalizer as well as the TV-LE, the performance is
significantly better than the TV-LE.

\begin{figure}[!t]
\centering
\includegraphics[width=16.0cm]{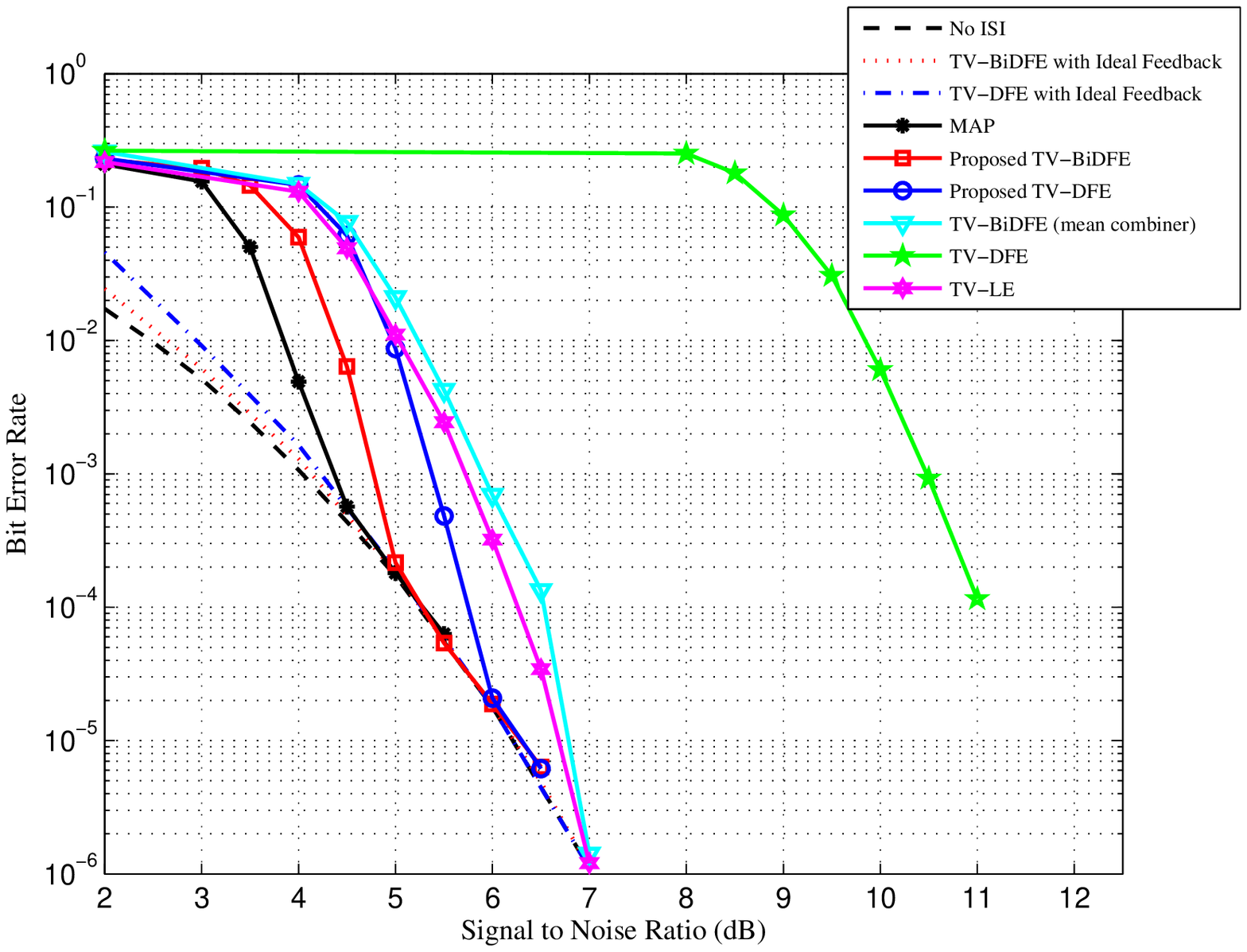}
\caption{BER Curve on the Channel $\mathbf{h_1}$ after 20 Iterations
with Time-varying Filters.}\label{fig:BER1_TV}
\end{figure}

Fig. \ref{fig:BER1_TV} shows the performance of several turbo
equalizers with time-varying filters after 20 iterations. TV-DFE
with the conventional LLR mapping shows poor performance but once
the proposed LLR generations are used (``Proposed TV-DFE"), the
DFE performance becomes clearly better than the TV-LE method of
\cite{TE02}, except at very high SNRs where all schemes other than
the conventional DFE perform comparably. The ``Proposed TV-BiDFE"
is considerably better than the TV-BiDFE based on the mean
combiner, approaching the performance of the MAP scheme.

\begin{figure}[!t]
\centering
\includegraphics[width=16.0cm]{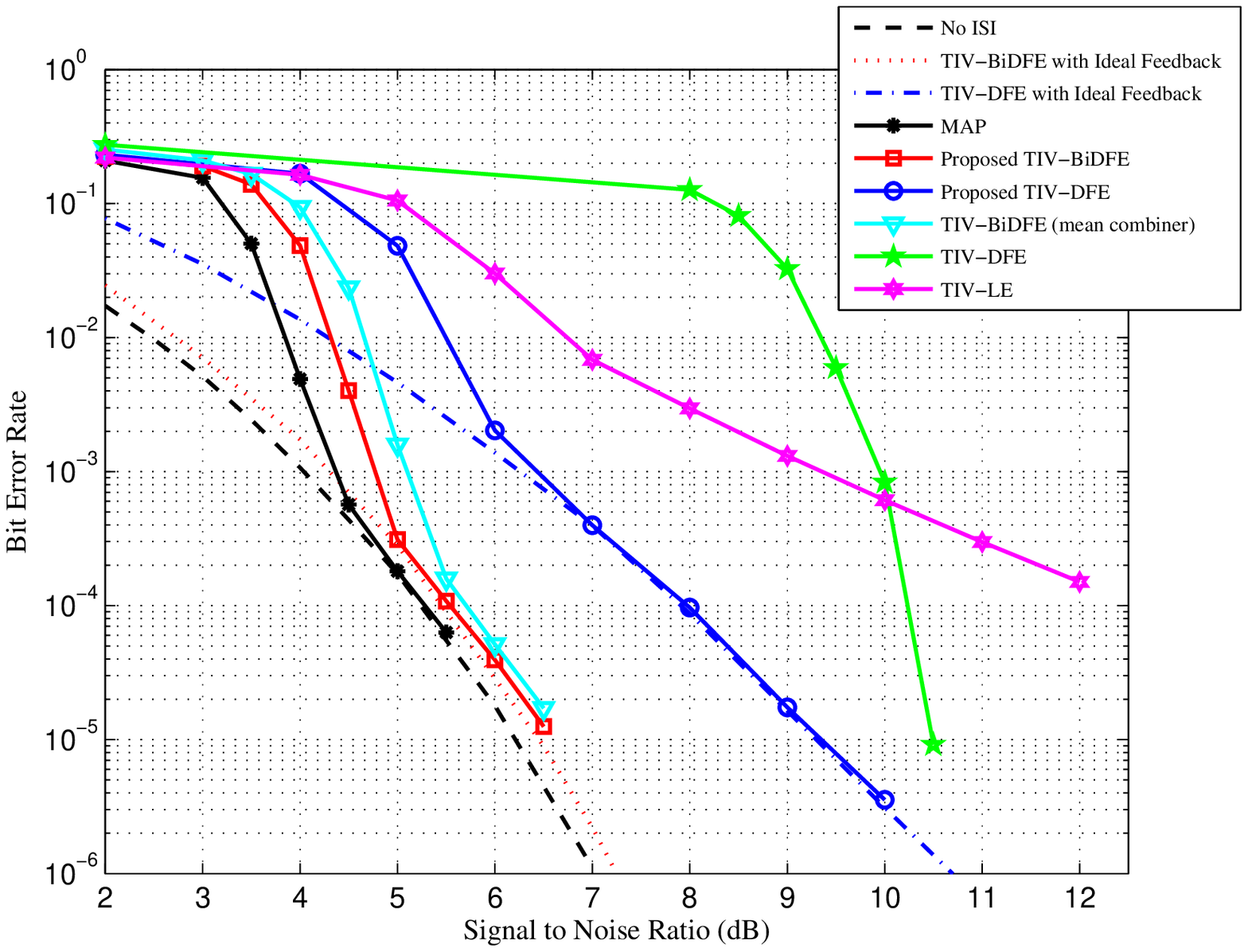}
\caption{BER Curve on the Channel $\mathbf{h_1}$ after 20 iterations
with Time-invariant Filters.}\label{fig:BER1_TIV}
\end{figure}

Fig. \ref{fig:BER1_TIV} shows the BER performance of
time-invariant-filter-based turbo equalizers. As the figure
indicates, the ``Proposed TIV-DFE" also shows superior performance
to the ``TIV-DFE". The performance of ``Proposed TIV-BiDFE" is
very close to the performance of the MAP equalizer while requiring
low computational complexity based on the use of time-invariant
filters. Also notice that both ``Proposed TIV-DFE" and ``Proposed
TIV-BiDFE" achieve decision-error-free performance at low BERs,
indicating the error propagation effect has been nearly eliminated
using the proposed LLR generation method. It is noteworthy that
the proposed BiDFE algorithm still provides near-optimal
performance even with the time-invariant filter taps. While the
TIV-BiDFE based on the existing mean combiner appears to perform
almost as well, the EXIT chart analysis to be discussed below
indicate that with a smaller number of turbo iterations, its
performance is distinctly inferior to the proposed TIV-BiDFE based
on the new combining method.

\begin{figure}[!t]
\centering
\includegraphics[width=16.0cm]{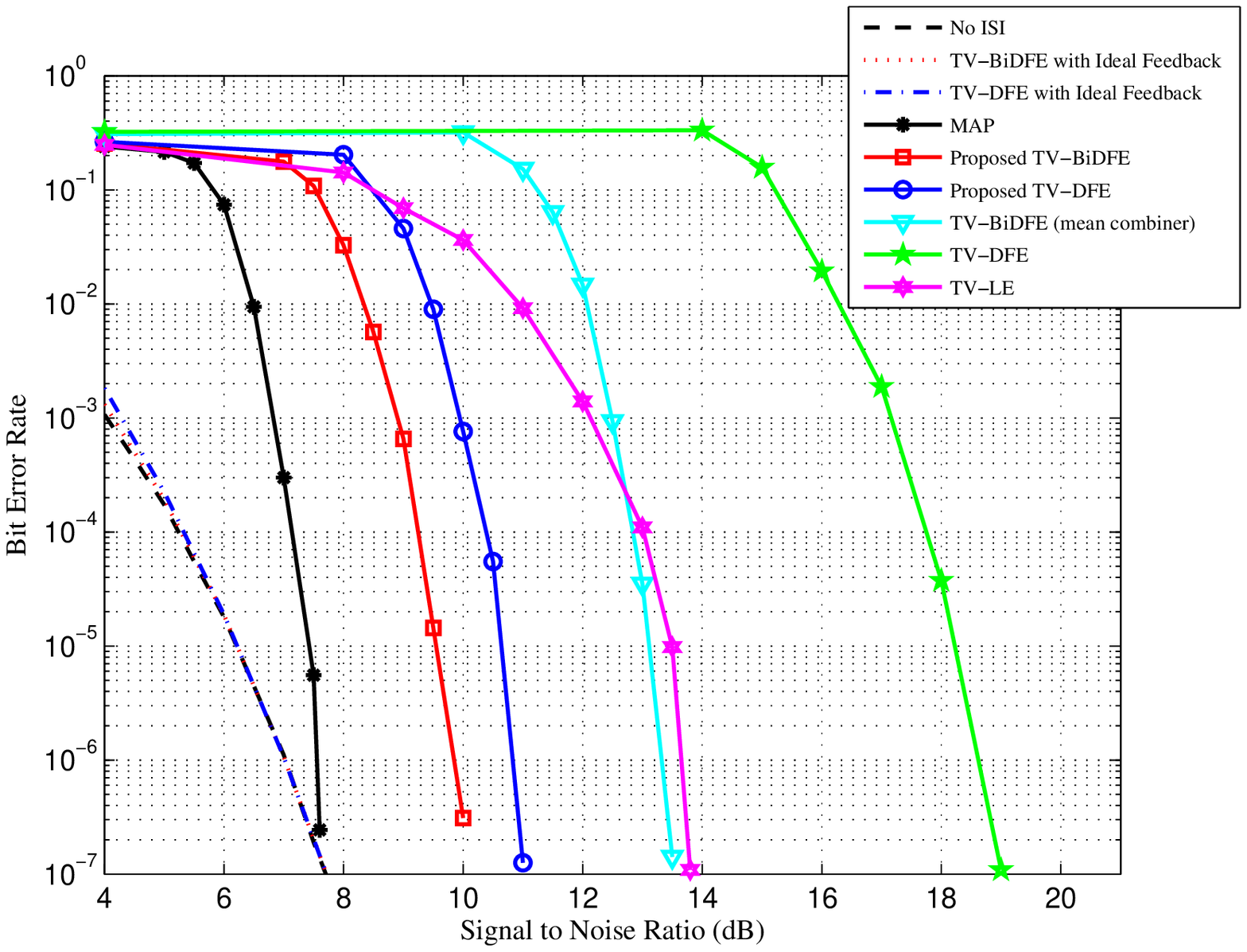}
\caption{BER Curve on the Channel $\mathbf{h_2}$ after 20 Iterations
with Time-varying Filters.}\label{fig:BER2_TV}
\end{figure}

\begin{figure}[!t]
\centering
\includegraphics[width=16.0cm]{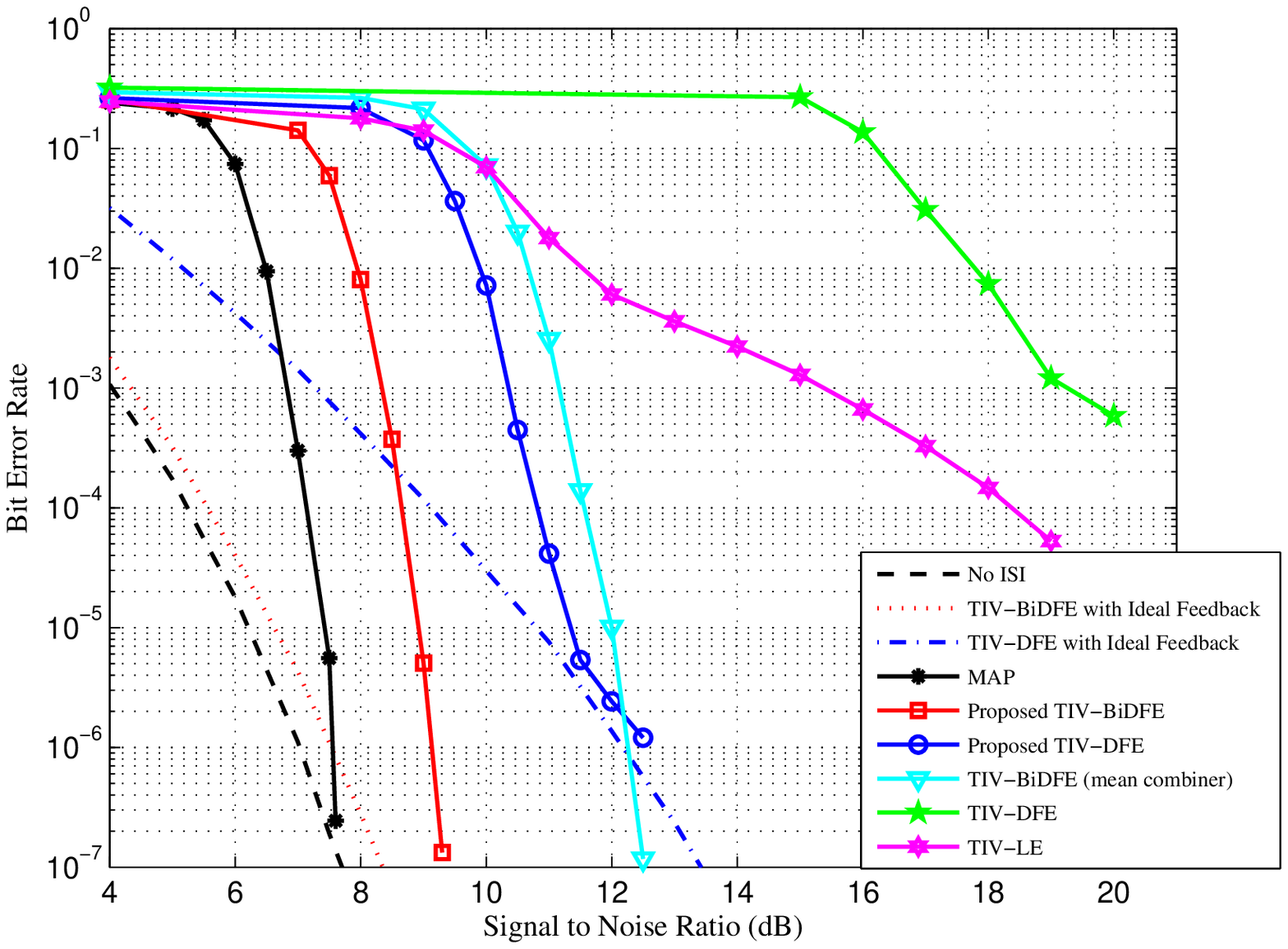}
\caption{BER Curve on the Channel $\mathbf{h_2}$ after 20 iterations
with Time-invariant Filters.}\label{fig:BER2_TIV}
\end{figure}

Figs. \ref{fig:BER2_TV} and \ref{fig:BER2_TIV} show a similar set
of simulation results now applied to the more severe ISI channel
$\mathbf{h_2}$. While all DFE-based schemes lag clearly behind the
BCJR-based scheme at the error rates simulated, the proposed BiDFE
scheme in both the time-varying and time-invariant filter cases
outperform the LE scheme by a significant margin. In fact, in this
severe channel the BER curve of the LE scheme, even with
time-varying filters, appears to diverge considerably from the ideal
no-ISI curve. Overall, the proposed BiDFE based on time-invariant
filter taps offer excellent performance-complexity trade-off.

\begin{figure}[!t]
\centering
\includegraphics[width=16.0cm]{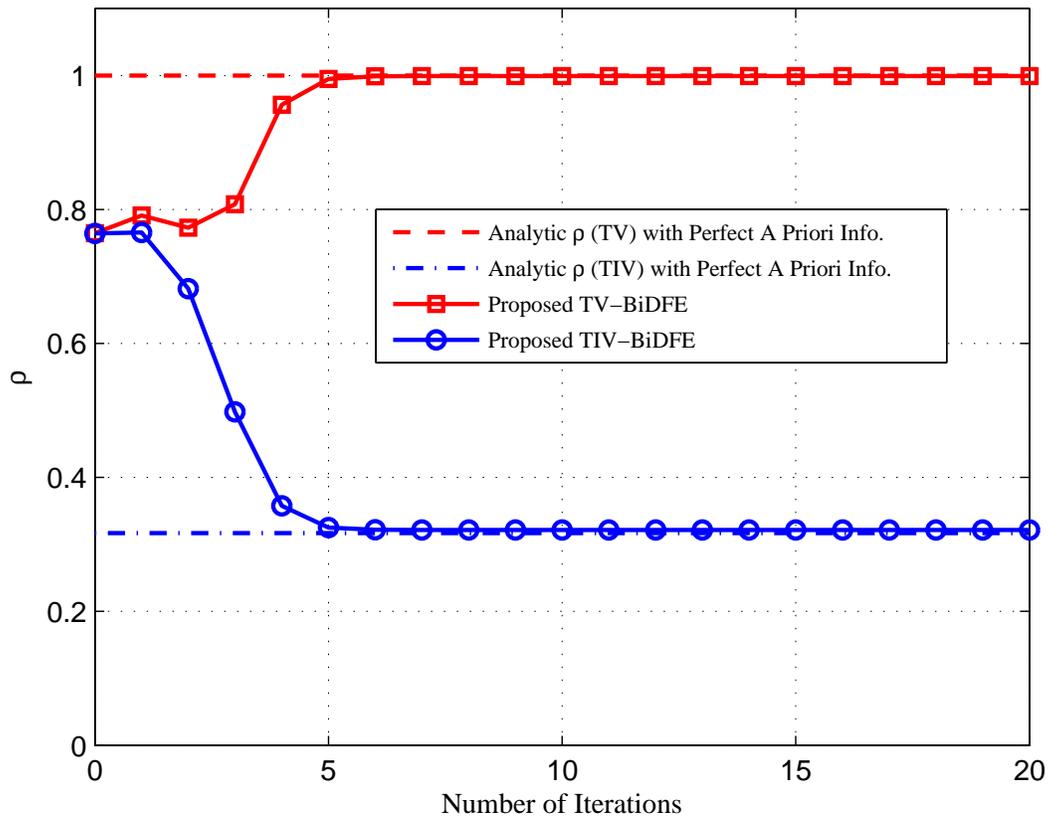}
\caption{Noise Correlation of ``Proposed BiDFE" on the Channel $\mathbf{h_1}$.}\label{fig:rho}
\end{figure}

The noise correlation in one block of coded data bits is described
in Fig. \ref{fig:rho}, at different iteration numbers at a 6 dB
SNR on $\mathbf{h_1}$. The correlation coefficient of ``Proposed
TV-BiDFE" goes to 1 as the number of iterations increases because
the \textit{a priori} information from the decoder becomes
reliable, and the time-varying filters in the normal and the
time-reversed DFEs produce essentially the same equalized output
sequences. This phenomenon of Fig. \ref{fig:rho} validates
(\ref{eq:tv_rho2}). On the other hand, the correlation coefficient
of ``Proposed TIV-BiDFE" actually decreases as the number of
iterations increases, and the noise correlation coefficient
converges to that of ``TIV-BiDFE with Ideal Feedback" or the
correlation coefficient of (\ref{eq:tiv_rho2}). This is because
the decision feedback errors disappear and the perfect \textit{a
priori} information is available from decoder. Note that the
filter coefficients in both DFEs do not change with the \textit{a
priori} information.

In general, it is quite difficult to analyse the iterative
equalization and decoding schemes. We rely on the oft-used
extrinsic information transfer (EXIT) chart of \cite{Brink01} to
develop insights into the convergence behaviour of the turbo
equalizers. The EXIT chart is a diagram demonstrating the mutual
information (MI) transfer characteristics of the two constituent
modules which exchange soft information. In the EXIT charts, the
behavior of the channel equalizer is described with its input and
output on the horizontal and vertical axis, respectively, while
the behavior of the decoder is described in opposite way. The pair
of EXIT chart curves typically defines a path for the MI
trajectory to move up during iterative processing of soft
information. The number of stairs that a given MI trajectory takes
to reach the highest value indicates the necessary number of
iterations toward convergence.

\begin{figure}[!t]
\centering
\includegraphics[width=16.0cm]{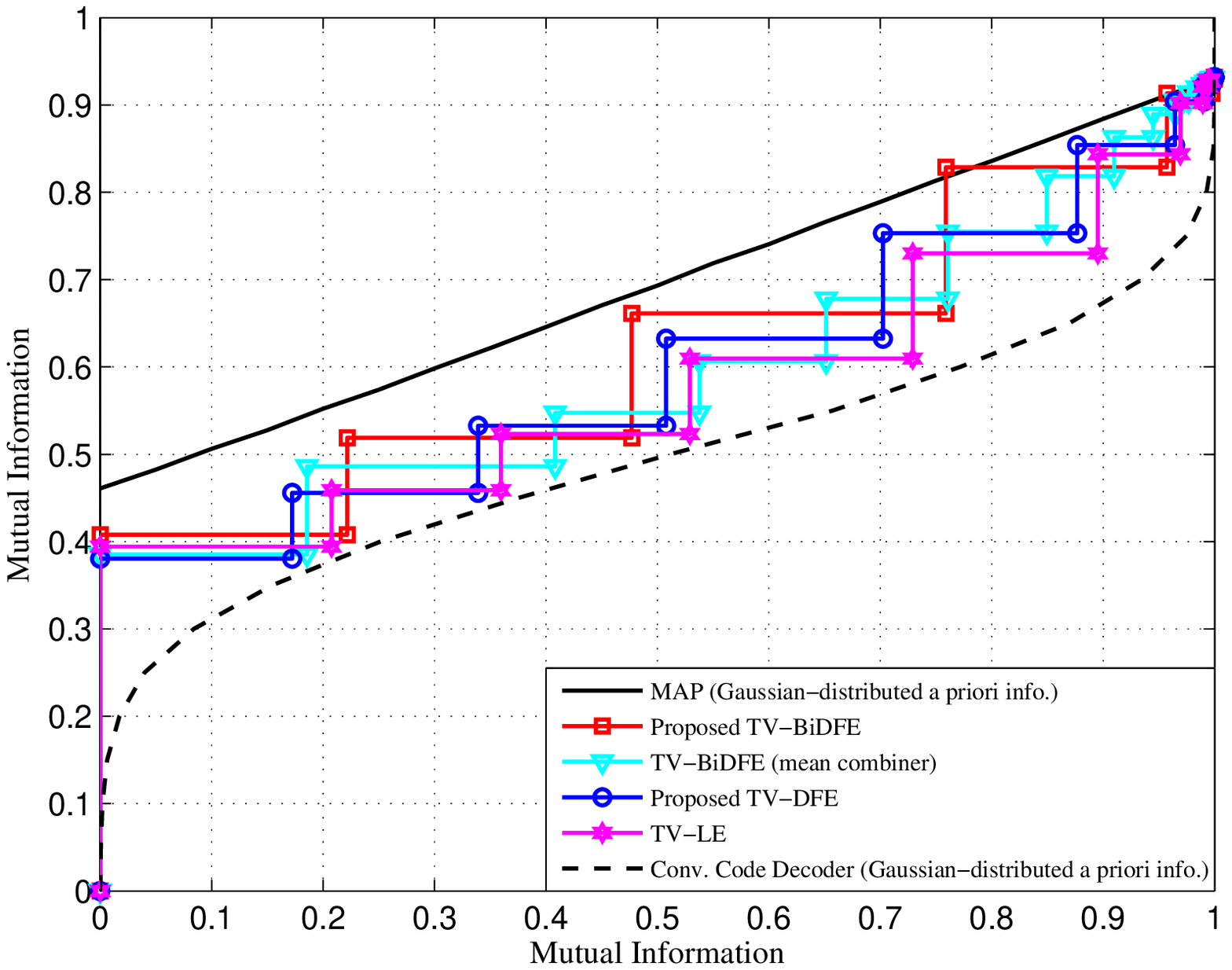}
\caption{EXIT Chart on the Channel $\mathbf{h_1}$ at a 6 dB with Time-varying Filters.}\label{fig:EXIT1_TV}
\end{figure}

\begin{figure}[!t]
\centering
\includegraphics[width=16.0cm]{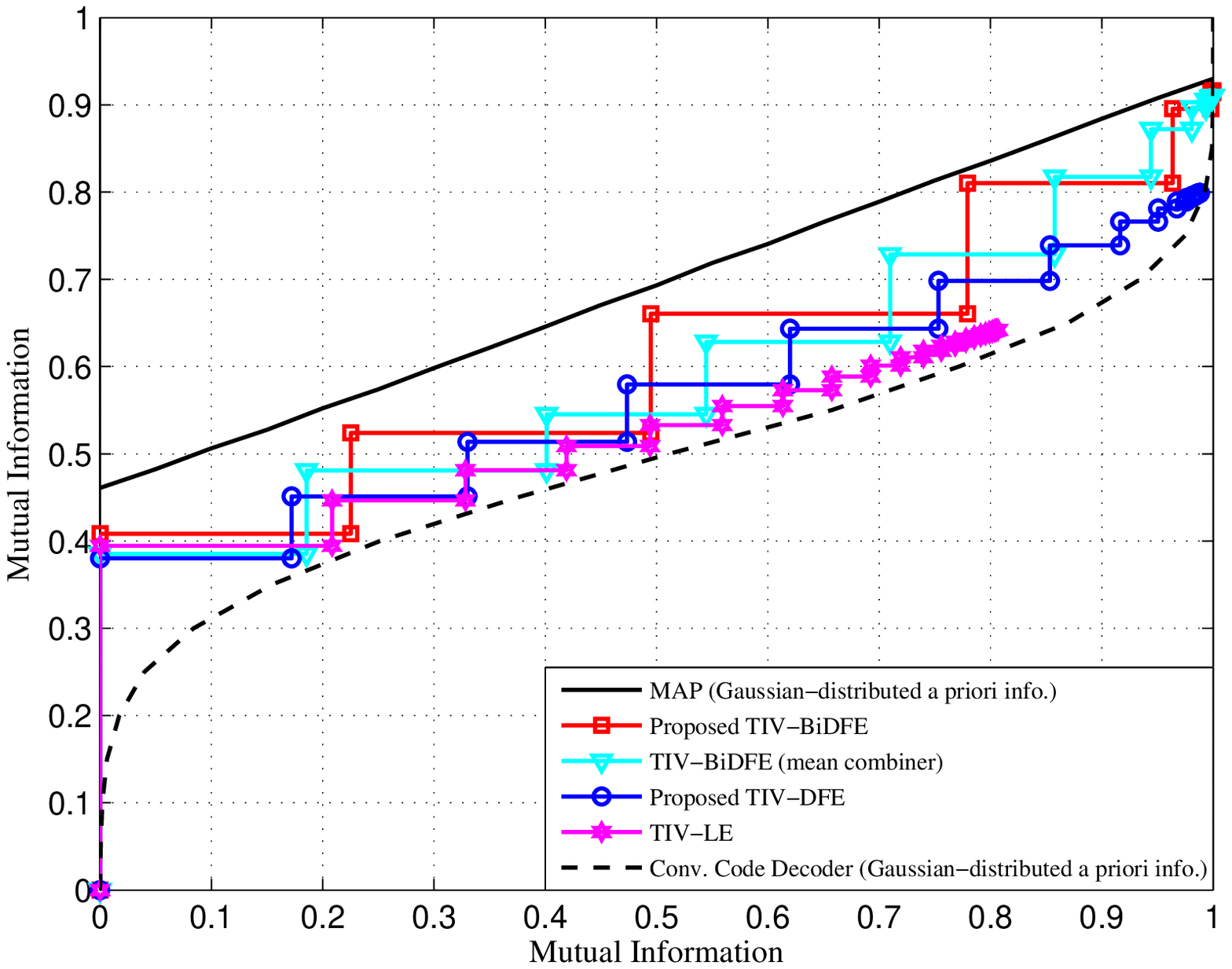}
\caption{EXIT Chart on the Channel $\mathbf{h_1}$ at a 6 dB with Time-invariant Filters.}\label{fig:EXIT1_TIV}
\end{figure}

\begin{figure}[!t]
\centering
\includegraphics[width=16.0cm]{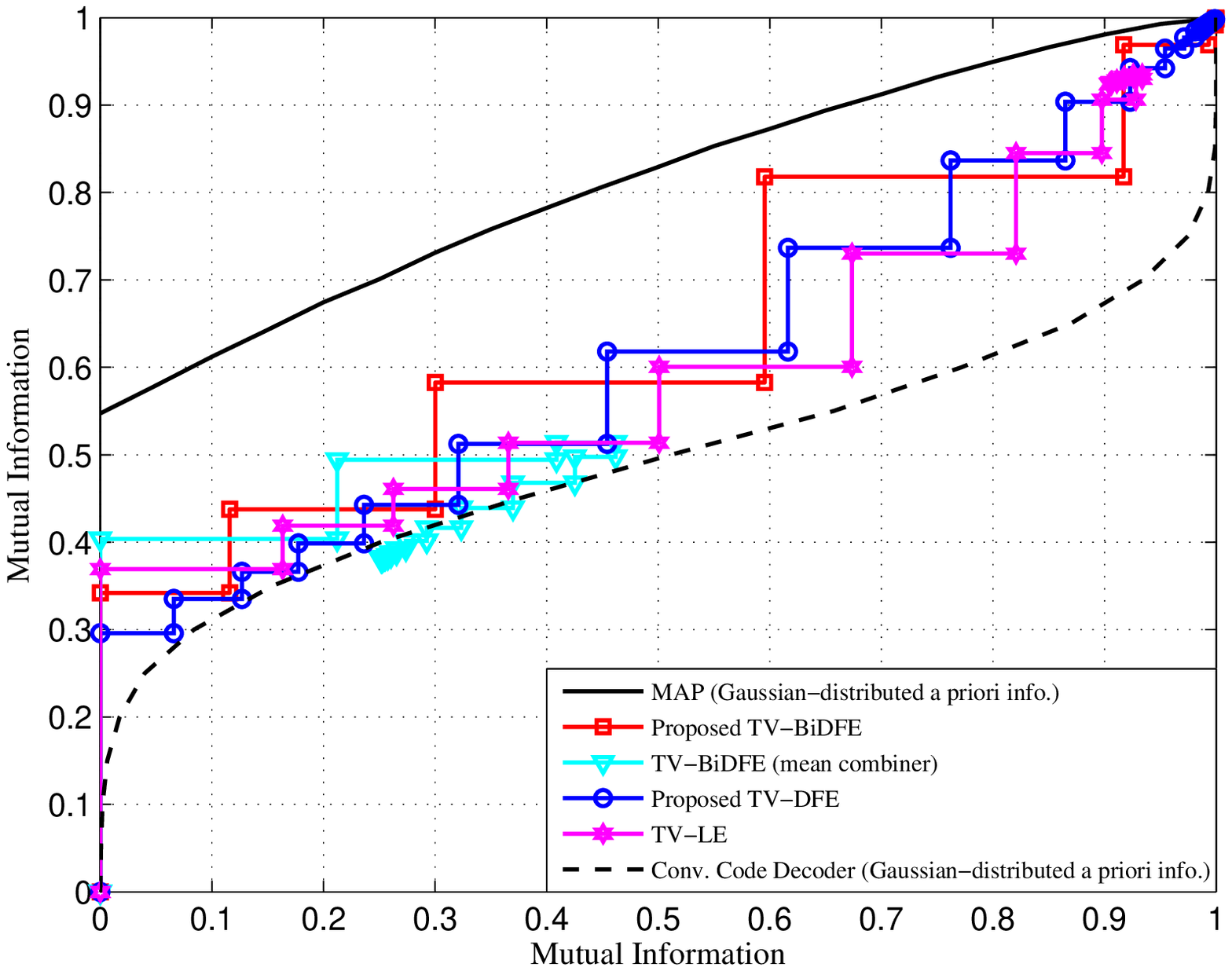}
\caption{EXIT Chart on the Channel $\mathbf{h_2}$ at a 10 dB with Time-varying Filters.}\label{fig:EXIT2_TV}
\end{figure}

\begin{figure}[!t]
\centering
\includegraphics[width=16.0cm]{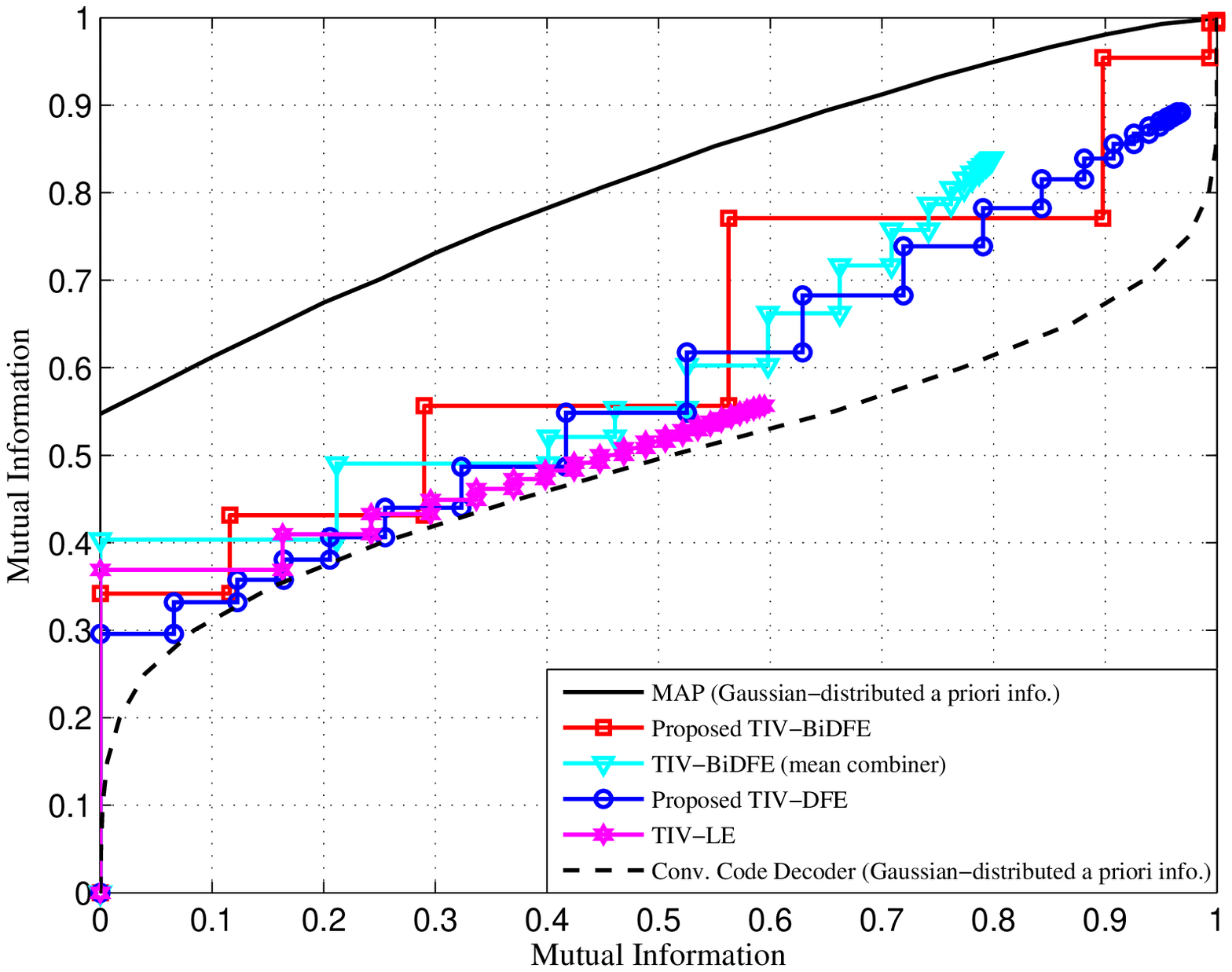}
\caption{EXIT Chart on the Channel $\mathbf{h_2}$ at a 10 dB with Time-invariant Filters.}\label{fig:EXIT2_TIV}
\end{figure}

Figs. \ref{fig:EXIT1_TV} and \ref{fig:EXIT2_TV} show the EXIT
chart corresponding to time-varying-filter-based equalizers for
$\mathbf{h_1}$ at a 6 dB SNR and $\mathbf{h_2}$ at a 10 dB SNR
while Figs. \ref{fig:EXIT1_TIV} and \ref{fig:EXIT2_TIV} show the
similar EXIT charts for time-invariant-filter-based schemes.
Although not shown here to avoid excessive cluttering, the
trajectories of ``TV-DFE" and ``TIV-DFE" move up for the first
couple of iterations, but then quickly fizzle out due to the
inadequate extrinsic LLR generations that cannot handle error
propagation. However, the trajectories of ``Proposed TV-DFE" and
``Proposed TIV-DFE" keep moving up as the number of iterations
increases, clearly indicating the advantage and effectiveness of
the proposed LLR generation method. However, the trajectory of
``Proposed TIV-DFE" at 6 dB or 10 dB does not reach the maximum
possible value since the filters do not fully exploit the
\textit{a priori} information from the decoder. The trajectories
of the ``Proposed TV-BiDFE" and ``Proposed TIV-BiDFE" indicate
that these schemes move from 0 bit of mutual information to 1 bit
with a less number of iteration runs than ``Proposed TV-DFE",
``Proposed TIV-DFE", ``TV-LE", or ``TIV-LE".

We notice, however, that the proposed BiDFE scheme requires more
iterations in achieving the full performance, relative to the MAP
equalizer (whose trajectory is not shown to avoid cluttering).
Nevertheless, the proposed BiDFE method offers a reasonable
tradeoff among complexity, performance, and latency.

\begin{figure}[!t] \centering
\includegraphics[width=16.0cm]{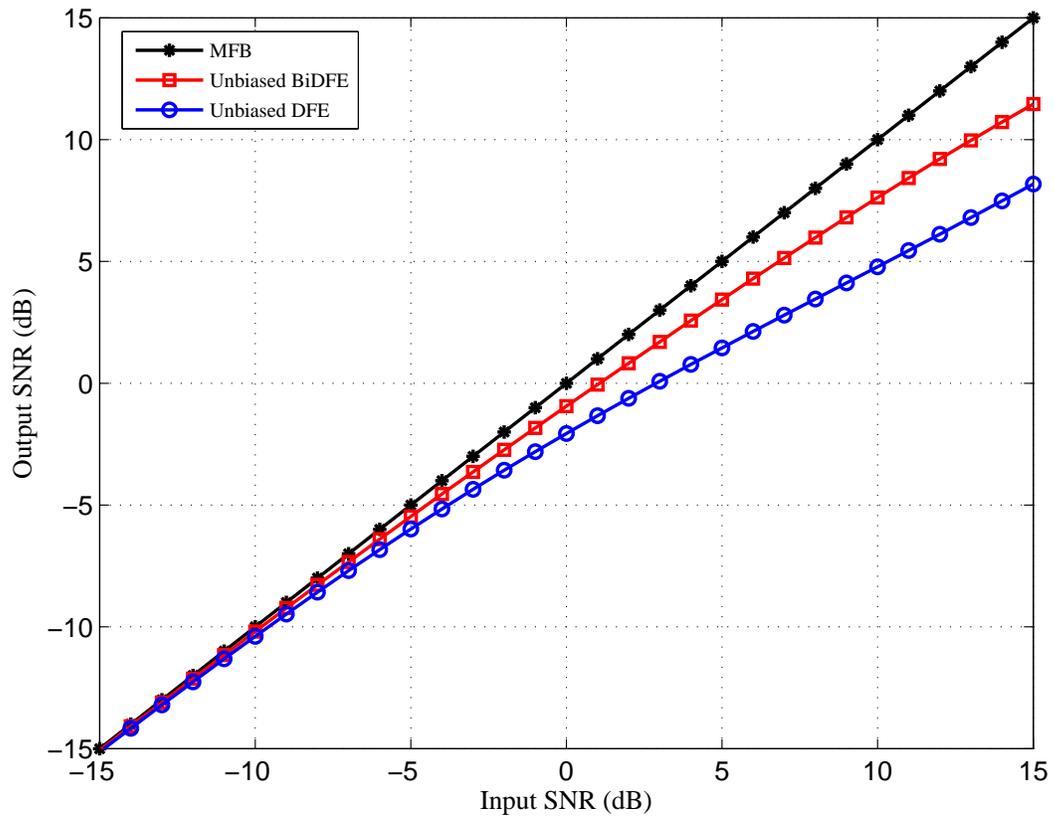}
\caption{SNR plot on the Channel $\mathbf{h_1}$.}\label{fig:SNR}
\end{figure}

Finally, Fig. \ref{fig:SNR} shows the SNR comparison at the output
of the unbiased DFE and BiDFE assuming ideal feedback on the
channel $\mathbf{h_1}$ when the \textit{a priori} information is
not available. As the figure shows, the output SNR of BiDFE is
considerably higher than the output SNR of DFE but with a certain
gap to the matched filter bound (MFB).

\section{Conclusion}\label{conclusion}
In this paper, we proposed new SISO DFE and BiDFE structures
well-suited to turbo equalization. The proposed LLR generation
designed to reduce error propagation indeed provides
decision-error-free performance in the DFE in turbo equalizer
setting. When further employing an LLR combining method that
estimates the correlation between the forward and backward DFE
outputs and whitens them, the resulting performance is remarkably
good given the simple structure of the BiDFE, relative to that of
the BCJR equalizer. The proposed LLR generation and combining
methods remain effective even when a time-invariance constraint is
imposed on the feedforward and feedback filters of the DFEs.
Overall, the proposed BiDFE method based on time-invariant filter
taps provides the excellent performance-complexity tradeoff for
severe ISI channels where the linear SISO equalizer fails to
operate adequately.

\end{document}